\documentclass[11pt]{article}
\usepackage{hyperref}  
\usepackage{tikz}
\usepackage{mdframed}
\usepackage{bbm}
\usepackage{adjustbox}
\usetikzlibrary{positioning}
\usepackage{epsfig,subfigure}
\usepackage{amssymb}
\usepackage{color}
\def\QED{\mbox{\rule[0pt]{1.5ex}{1.5ex}}}

\usepackage{graphicx}

\usepackage{amssymb}
\usepackage{amsmath,amsfonts,amssymb}
\usepackage{verbatim}
\usepackage{stfloats}
\usepackage[bookmarks=false]{}
\usepackage{cite}

\newtheorem{theorem}{Theorem}

\newtheorem{definition}{Definition}
\newtheorem{lemma}{Lemma}

\usepackage[normalem]{ulem}

\newcommand\blfootnote[1]{%
  \begingroup
  \renewcommand\thefootnote{}\footnote{#1}%
  \addtocounter{footnote}{-1}%
  \endgroup
}
\def\cb{{\mbox{\tiny\it CB}}}

\begin{document}
\date{}
\title{
On the Capacity of Computation Broadcast
}
\author{ 
\normalsize Hua Sun and Syed A. Jafar \\
}

\maketitle

\blfootnote{Hua Sun (email: hua.sun@unt.edu) is with the Department of Electrical Engineering at the University of North Texas. Syed A. Jafar (email: syed@uci.edu) is with the Center of Pervasive Communications and Computing (CPCC) in the Department of Electrical Engineering and Computer Science (EECS) at the University of California Irvine. This work was presented in part at ICC 2019. }

\begin{abstract}
The two-user computation broadcast problem is introduced as the setting where User $1$ wants message $W_1$ and has side-information $W_1'$, User $2$ wants message $W_2$ and has side-information $W_2'$, and $(W_1, W_1', W_2, W_2')$ may have arbitrary dependencies. The rate of a computation broadcast scheme is defined as the ratio $H(W_1,W_2)/H(S)$, where $S$ is the information broadcast to both users to simultaneously satisfy their demands. The supremum of achievable rates is called the capacity of computation broadcast $C_\cb$. It is shown that $C_\cb\leq H(W_1,W_2)/\left[H(W_1|W_1')+H(W_2|W_2')-\min\Big(I(W_1; W_2, W_2'|W_1'), I(W_2; W_1, W_1'|W_2')\Big)\right]$. For the linear computation broadcast  
problem, where $W_1, W_1', W_2, W_2'$ are comprised of arbitrary linear combinations of a basis set of independent symbols,  the bound is shown to be tight. For non-linear computation broadcast, it is shown that this bound is not tight in general. Examples are provided to prove that different instances of computation broadcast that have the same entropic structure, i.e., the same entropy for all subsets of $\{W_1,W_1',W_2,W_2'\}$, can have different capacities. Thus, extra-entropic structure matters even for two-user computation broadcast. The significance of extra-entropic structure is further explored through a class of non-linear computation broadcast problems where the extremal values of capacity are shown to correspond to minimally and maximally structured problems within that class.  
\end{abstract}

\newpage
\allowdisplaybreaks
\section{Introduction}
In the modern era of data science,  machine learning and internet of things, communication networks are increasingly used for distributed computing applications, where multiple parties process and exchange information for various computational tasks \cite{Lee_Lam_Pedarsani_Papailiopoulos_Ramchandran, Li_Maddah_Yu_Avestimehr, Yu_Raviv_So_Avestimehr, Tandon_Lei_Dimakis_Karampatziakis, Dutta_Cadambe_Grover, Sun_Jafar_PC, Braverman, Lee_Shraibman}. The changing paradigm brings with it new challenges in network information theory. Distinctive aspects of these computational communication networks include strong dependencies between information flows and an abundance of side-information. With a few notable exceptions such as \cite{Slepian_Wolf, Cover_Gamal_Salehi, Han_Costa, Salehi_Kurtas, Pradhan_Choi_Ramchandran, Tuncel_SW, Gunduz_Erkip_Goldsmith_Poor, Liu_Chen, Liu_Gunduz_Goldsmith_Poor}, the communication network models most commonly studied in information theory, in various elemental forms ranging from multiple access and broadcast to relay and interference networks, with and without side-information, tend to focus on settings with  \emph{independent} messages. Yet, the  shared mission across network nodes in distributed computing applications necessarily creates significant dependencies, not only among  message flows, but also in the side-information available to each node based on its history of prior computations. Within these dependencies lies the potential for  further innovations in communication and computing. A fundamental understanding of this potential requires the machinery of network information theory, but with a renewed focus on information dependencies and side-information. As a step in this direction, in this work we introduce the problem of computation broadcast.

While in this work we restrict our attention to $K=2$ users, in general we envision the computation broadcast (CB) problem as comprised of $K$ users (receivers) who desire messages $W_1, W_2, \cdots, W_K$, and have prior side-information $W_1', W_2', \cdots, W_K'$, respectively.  A centralized transmitter with full knowledge of $(W_k, W_k', k\in[K])$  broadcasts the same  information $S$ to all receivers in order to simultaneously deliver their desired messages. The salient feature of computation broadcast is the dependence among $(W_k, W_k', k\in[K])$ modeled by their joint distribution, which may be arbitrary.

The rate of computation broadcast is defined as, $R = H(W_1,\cdots,W_K)/H(S)$, i.e., ratio of the total number of bits of all desired messages to the number of bits of  broadcast information $S$ that  satisfies all demands. 
The supremum of achievable rates is called the capacity of computation broadcast, $C_\cb$. The goal is to characterize $C_\cb$.

The computation broadcast problem may be seen as a  generalization of the index coding problem \cite{Birk_Kol, Yossef_Birk_Jayram_Kol} that allows arbitrary dependencies among desired messages and side-informations. Prior works in this direction include   \cite{Lee_Dimakis_Heath, Miyake_Muramatsu}. Reference \cite{Lee_Dimakis_Heath}  restricts the messages to be independent and requires each side-information to be a linear combination of message symbols, which is a special case of computation broadcast. The problem formulation of \cite{Miyake_Muramatsu} allows the messages to be arbitrarily correlated while the side-informations are comprised of message symbols, which is another special case of computation broadcast. Also, when we have $K=2$ users and $W_1' = W_2$ and $W_2' = W_1$, the computation broadcast problem reduces to the classic butterfly network problem with possibly correlated sources \cite{Li_etal, Gohari_Amin_Yang_Jaggi}.

The dependence between desired messages and side-informations imparts a unique structural aspect to the computation broadcast problem that makes it highly non-trivial. Structure has long been recognized as both the boon and bane of network information theory \cite{Korner_Marton_sum, Philosof_Zamir, Pradhan_Choi_Ramchandran_Graph, Nazer_Gastpar_Compute, Zamir_Lattice}.  When optimally exploited, structure can have tremendous benefits in multiterminal settings, a fact underscored by recurring observations ranging from Korner and Marton's computation work in \cite{Korner_Marton_sum} to the recent burst of activity in interference alignment  \cite{Cadambe_Jafar_int, Jafar_FnT}. On the other hand, the random coding arguments that are the staple of classical information theory, tend to fall short when structural concerns take center stage,  and less tractable combinatorial alternatives are required. Structure itself is a nebulous concept that has thus far defied a precise understanding. Somewhat surprisingly, these deeper themes resurface even in the basic $2$ user setting explored in this work. On the downside this potentially makes even the $2$ user computation broadcast problem  intractable in general. On the upside, the $2$ user computation broadcast presents one of the simplest arenas to face these challenges that are of tremendous theoretical and practical significance.

Our contributions in this paper are summarized as follows. We start with a general converse bound for the capacity of $2$ user computation broadcast, $$C_\cb\leq H(W_1,W_2)/\left[H(W_1\mid W_1')+H(W_2\mid W_2')-\min\Big(I(W_1;W_2,W_2'\mid W_1'), I(W_2;W_1,W_1'\mid W_2')\Big)\right].$$ When the dependency is linear, i.e., when $W_1, W_1', W_2, W_2'$ are comprised of arbitrary linear combinations of a basis set of independent symbols, then this bound is shown to be tight. However, in general the bound is not tight, and the  structure of the dependence between $W_1, W_1', W_2, W_2'$, becomes important. Recall that the dependence is completely described by the joint distribution of $(W_1, W_1', W_2, W_2')$ which can be arbitrary. Some of this structure can be captured through entropic constraints, i.e., the joint entropies of all subsets of $(W_1, W_2, W_1', W_2')$. One might optimistically expect that only this entropic structure would be essential to the problem, and furthermore that Shannon information inequalities might suffice to characterize the optimal $H(S)$. However, as it turns out on both counts the optimism is invalidated. Specifically, we show two instances of computation broadcast that have the same entropic description, yet different capacity characterizations. Evidently, extra-entropic structure matters even for $2$-user computation broadcast. In order to further understand the significance of such extra-entropic structure, we explore a class of computation broadcast problems called `matching' problems where, conditioned on each realization of the independent side-informations $W_1', W_2'$, there is a perfect matching between the possible realizations of desired messages $W_1, W_2$. For this class of problems we identify  upper and lower bounds on capacity. The bounds provide insights into certain types of extremal structures that are either beneficial or detrimental to capacity. The beneficial extremes are found to be maximally structured and for these settings the capacity upper bound is shown to be tight. Conversely, the detrimental extremes are found to be minimally structured and for these settings the capacity lower bound is shown to be tight.  Remarkably, linear dependencies are maximally structured, while random coding solutions are asymptotically optimal for minimally structured settings in the limit of large alphabet sizes.

{\it Notation: For a positive integer $m$, we use the notation $[m]=\{1,2,\cdots, m\}$. Bold symbols are used to represent matrices.}

\section{Problem Statement and Preliminaries}\label{sec:model}
Define random variables  $(w_1, w_1', w_2, w_2')\in \mathcal{W}_1\times\mathcal{W}_1'\times\mathcal{W}_2\times \mathcal{W}_2'$, drawn according to an arbitrary joint distribution $P_{w_1, w_1', w_2, w_2'}$. All $4$ alphabet sets are discrete with  finite cardinality bounded by $2^{\ell_{\max}}<\infty$, i.e., it takes no more than a finite number $(\ell_{\max})$  of bits to perfectly specify any $w_i, w_i'$, $i\in\{1,2\}$. 

\subsection{Complete (Structural) Formulation}\label{sec:structure}
The complete formulation of the computation broadcast problem is presented as follows.
\begin{eqnarray}
R^*_L\triangleq \sup_{P_{S\mid W_1,W_2,W_1',W_2'}} \frac{H(W_1, W_2)}{H(S)} &&\notag\\
\mbox{such that }~~~~~~
H(W_1\mid W_1', S)&=&0 \label{eq:dec1}\\
H(W_2\mid W_2', S)&=&0 \label{eq:dec2} \\
{[(W_1(l), W_1'(l), W_2(l), W_2'(l))]}_{l=1}^L &\stackrel{\mbox{\tiny i.i.d.}}{\sim}& P_{w_1, w_1', w_2, w_2'} \label{eq:structure}
\end{eqnarray}
As indicated in \eqref{eq:structure},  $W_1, W_2, W_1'$, $W_2'$ denote $L$ length extensions of $w_1, w_1', w_2, w_2'$, i.e., $W_1, W_1', W_2$, $W_2'$ are sequences of length $L$, such that the sequence of tuples $[(W_1(l), W_1'(l), W_2(l), W_2'(l))]_{l=1}^L$ is  produced i.i.d. according to $P_{w_1, w_1', w_2, w_2'}$. Because the structure of the problem is completely captured in \eqref{eq:structure}, we refer to this problem formulation as the complete, or structural formulation.  $L$ is called the block length. $H(S)$ is the expected  amount of broadcast information. Condition (\ref{eq:dec1}) is the decoding constraint of User 1, i.e., after receiving the broadcast information $S$, User 1 is able to decode his desired message $W_1$  with the help of  the side-information $W_1'$, with zero probability of error. Similarly, condition (\ref{eq:dec2}) is the decoding constraint of User 2.  Note that $H(W_1,W_2)$ is already specified by the problem statement, so maximizing $R_L^*$ is the same as minimizing the broadcast cost, $H(S)$. The ratio $H(W_1,W_2)/H(S)$ for a computation broadcast scheme is called its achievable rate. $R_L^*$ is the supremum of achievable rates for a given block length $L$. The supremum of $R_L^*$ across all $L\in\mathbb{N}$, is called the capacity of computation broadcast,
\begin{align}
~~~~~~~~~~C_\cb &\triangleq \sup_{L\in\mathbb{N}} {R^*_L}.
\end{align}

\subsection{Relaxed (Entropic) Formulation}\label{sec:entropic}
Recall that the structure of the dependence between message and side-information random variables is defined by Condition \eqref{eq:structure}. Some of this structure can be captured in terms of  the entropies of all subsets of $\{w_1,w_2,w_1',w_2'\}$. Limited to just these entropic constraints we obtain the following relaxed problem formulation.
\begin{align}
\overline{R}_L^*\triangleq \sup_{\bar{P}_{W_1, W_2, W_1', W_2', S}} \frac{H(W_1, W_2)}{H(S)}  &&\notag\\
\mbox{such that }~~~~~~
H(W_1\mid W_1', S)&=0 \label{eq:dec11}\\
H(W_2\mid W_2', S)&=0 \label{eq:dec22} \\
H(W_*)&=LH(w_*), &&\forall W_*\subset\{W_1,W_2,W_1',W_2'\}
 \label{eq:Hstructure}\\
w_1,w_2,w_1',w_2'&\sim P_{w_1,w_2,w_1',w_2'}
\end{align}
where $w_*$ is obtained by replacing upper case $W$ with lower case $w$ in $W_*$. For example, if $W_*=(W_1,W_2')$, then $w_*=(w_1,w_2')$. {\color{black}  Note that $(W_1, W_2, W_1', W_2')$ are arbitrary random variables that only need to satisfy the same entropic constraints as the $L$-length extensions of $(w_1, w_2, w_1', w_2')$, according to \eqref{eq:Hstructure}. In particular, it is no longer necessary for $(W_1, W_2, W_1', W_2')$ to have the same distribution as $(w_1, w_2, w_1', w_2')$, even for $L=1$. Furthermore, since the entropic region is a cone \cite{ZY_Nonshannon}, we must have $\overline{R}_L^* =  \overline{R}_1^*$, where $\overline{R}_1^*$ is the value of $\overline{R}_L^*$ for $L=1$. Since $L$ is a trivial scaling factor, let us fix $L=1$, and define
\begin{align}
\overline{C}_\cb &\triangleq \sup_{L\in\mathbb{N}} \overline{R}_L^* = \overline{R}_1^* 
\end{align}
$\overline{C}_\cb$ is of interest mainly for two reasons. First, because it serves as a bound for $C_\cb$, i.e., 
\begin{align}
 C_\cb \leq \overline{C}_\cb. \label{eq:deltastructure}
\end{align}
This is true because all the entropic constraints  \eqref{eq:Hstructure} are implied by Condition \eqref{eq:structure}, so we must have ${R}^*_L \leq \overline{R}_L^*$ which in turn implies that $C_\cb \leq \overline{C}_\cb$. } The second reason is that  the tightness of the bound \eqref{eq:deltastructure} reveals the extent to which capacity is determined by structural constraints that are not captured by the entropic formulation. This extra-entropic structure may be a topic of interest by itself.

\subsection{Equivalence of zero-error and $\epsilon$-error}
While we consider the zero-error capacity formulation, it turns out that for the computation broadcast problem, it is not difficult to prove that zero-error capacity is the same as $\epsilon$-error capacity, as stated in the following theorem. For this theorem we use the specialized notation $C_{\cb}^{\mbox{\tiny \it 0}}$ to denote zero-error capacity, and $C_\cb^\epsilon$ to denote $\epsilon$-error capacity.
\begin{theorem}
For the computation broadcast problem,  zero error capacity, $C_{\cb}^{\mbox{\tiny 0}}$, is equal to  $\epsilon$-error capacity, $C_{\cb}^\epsilon$ .
\end{theorem}

{\it Proof:} 
Since the $\epsilon$-error capacity is $C_\cb^\epsilon$, for any arbitrarily small $\delta>0$, there must exist an $\epsilon$-error scheme that achieves rate $R_\epsilon=C_\cb^\epsilon-\delta$, so that broadcasting $L{\color{black}H(w_1,w_2)}/R_\epsilon$ bits is sufficient to satisfy both users' demands with probability at least $1-\epsilon$, and $\epsilon \rightarrow 0$ as $L \rightarrow \infty$. Since the encoder knows all messages, side-informations and decoding functions, it also knows when either decoding function will produce an erroneous output. In those cases, the encoder can simply use uncoded broadcast to send both messages using no more than $2L\ell_{\max}$ bits. One extra bit, say the first bit, is used to indicate when uncoded transmission takes place. Thus we have a zero-error scheme, and the rate achieved is
\begin{eqnarray}
\frac{L{\color{black}H(w_1,w_2)}}{(1-\epsilon)(L{\color{black}H(w_1,w_2)}/R_\epsilon) + \epsilon (2L\ell_{\max})+1} \overset{L\rightarrow \infty}{\longrightarrow} R_\epsilon
\end{eqnarray}
 Since the rate $R_\epsilon=C_\cb^\epsilon-\delta$ is asymptotically achievable with zero probability of error for any $\delta>0$, the zero error capacity $C_{\cb}^{\mbox{\tiny 0}}$, which is the supremum of  rates achievable with zero-error, cannot be less than $C_\cb^\epsilon$. At the same time, $C_{\cb}^{\mbox{\tiny \it 0}}$ cannot be more than $C_\cb^\epsilon$ because allowing $\epsilon$ decoding error cannot hurt. Therefore, we must have $C_{\cb}^{\mbox{\tiny\it 0}}=C_\cb^\epsilon$.\hfill\QED

\interfootnotelinepenalty=0
\subsection{Introductory Examples} 
\subsubsection{Example 1: The Butterfly Network}
For our first example, consider $(w_1,w_2,w_1',w_2')=(A,B,B,A)$, where $A,B$ are i.i.d. uniform over some finite field $\mathbb{F}_q$. This is the  butterfly network that is one of the most recognizable settings for network coding and index coding. The solution is also well known. The capacity is {\color{black}$2$} and is achieved by broadcasting $S=A+B$ (the addition is in $\mathbb{F}_q$) to simultaneously satisfy both users' demands. The example can be generalized  to $(w_1, w_2, w_1', w_2')$ where $w_1$ is a function of $w_2'$ and $w_2$ is a function of $w_1'$. In this case,  we need a codeword of $H(w_1\mid w_1')$ bits to satisfy User $1$, corresponding to the bin index when  $w_1$ is binned according to Slepian-Wolf coding (does not need the knowledge of $w_1'$ at the encoder). These bits are known to User $2$, because User $2$ knows the binning function as well as $w_2'$, and $w_1$ is a function of $w_2'$. Similarly, we need $H(w_2\mid w_2')$ bits to satisfy User $2$, and these bits are known to User $1$. Therefore, we can choose $S$ as the bitwise XOR of the two codewords (padding with additional zeros if needed so we have equal number of bits for both codes), which satisfies both users' demands. So the capacity for this case is {\color{black}$C_\cb=\frac{H(w_1,w_2)}{\max(H(w_1\mid w_1'), H(w_2\mid w_2'))}$.}

\subsubsection{Example 2: A Minimal Linear Dependence Setting}
Consider $w_1, w_2, w_1', w_2'$, all in $\mathbb{F}_q$, with a `minimal' dependence among them in the sense that any  three of these four random variables are independent and uniform, while the dependence arises due to the constraint $w_1+w_2+w_1'+w_2'=0$. In this case, the capacity is still ${\color{black}2}$, and it is achieved by broadcasting $S=w_1+w_1'$, which simultaneously satisfies both users. This example is inspired by a general capacity achieving scheme for linear computation broadcast problems that is developed in this work.

\subsubsection{Example 3: A Binary AND/OR Problem}
For our third example, let us consider a non-linear computation broadcast problem, where we have $(w_1, w_2, w_1', w_2')$ $=$ $(A\lor B, A\land B, A, B)$, and $A, B$ are independent uniform binary random variables. The notations $\lor, \land$ represent the logical OR and AND operations, respectively. Thus, User $1$ knows $A$ and wants $A\lor B$, while User $2$ knows $B$ and wants $A\land B$.

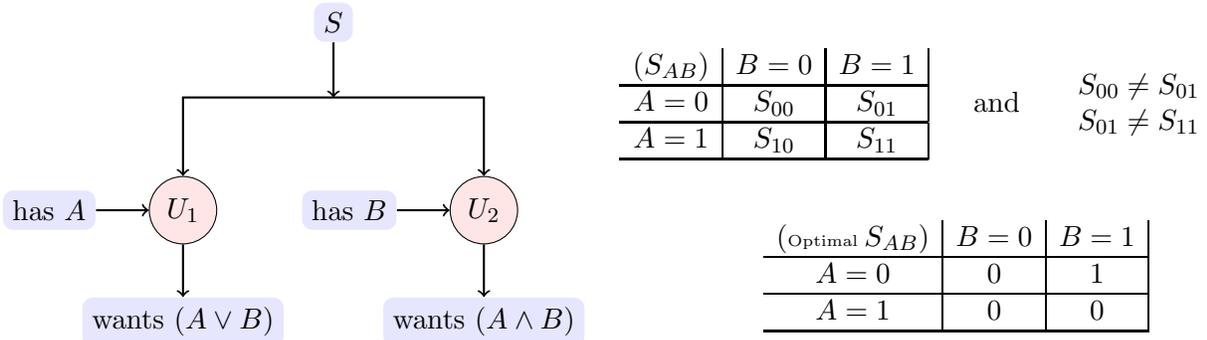
\begin{figure}[!h]
\begin{adjustbox}{valign=c,minipage={.45\textwidth}}
\begin{tikzpicture}

\draw (-2,-0.5) node[ rounded corners, fill=blue!10] (X)  { $S$};
\node[circle, draw=black, fill=red!10] (U1) at (-4,-3)  { $U_1$};
\node[circle, draw=black, fill=red!10] (U2) at (0,-3)  { $U_2$};
\node[left=0.7cm of U1, align=center,  rounded corners, fill=blue!10](S1)  {has $A$};
\node[below=0.7cm of U1, align=center, rounded corners, fill=blue!10](W1)  {wants $(A \lor B)$};
\node[left=0.7cm of U2, align=center,  rounded corners, fill=blue!10](S2)  {has $B$};
\node[below=0.7cm of U2, align=center,  rounded corners, fill=blue!10](W2)  {wants  $(A\land B)$};
\draw[thick, ->] (X) -- node [near start, right](A) {} (-2, -1.5);
\draw[thick, ->] (-2,-1.5) -| node [midway, above=0.2cm] {}(U1);
\draw[thick, ->] (-2,-1.5) -| node [midway, above=0.2cm] {} (U2);
\draw[thick, ->] (U1) -- (W1);
\draw[thick, ->] (U2) -- (W2);
\draw[thick, ->] (S1) -- (U1);
\draw[thick, ->] (S2) -- (U2);
\end{tikzpicture}
\end{adjustbox}~
\begin{adjustbox}{valign=c,minipage={.55\textwidth}}
\begin{align*}
\begin{array}{c|c|c|}
(S_{AB})&B=0&B=1\\ \hline
A=0&S_{00}&S_{01}\\ \hline
A=1&S_{10}&S_{11}\\ \hline
\end{array}&& \mbox{and} &&
\begin{array}{c}
S_{00}\neq S_{01}\\
S_{01}\neq S_{11}
\end{array}
\end{align*}
\begin{eqnarray*}
\begin{array}{c}
\\
\end{array}
&&
\begin{array}{c|c|c|}
(\mbox{\tiny Optimal } S_{AB})&B=0&B=1\\ \hline
A=0&0&1\\ \hline
A=1&0&0\\ \hline
\end{array}
\end{eqnarray*}
\end{adjustbox}
\caption{\footnotesize Toy example where User $1$ has side-information $A$ and wants to receive $(A\lor B)$ while User $2$ has side-information $B$ and wants $(A\land B)$. The optimal solution  broadcasts only $0.5$ bits/symbol to simultaneously satisfy both users' demands.}
\label{fig:ANDOR}
\end{figure}

Note that the desired message and available side-information are not independent. Also note that in order to satisfy User $1$ alone, we need at least $H(A\lor B|A)=0.5$ bits/symbol. Similarly, in order to satisfy User $2$ alone, we need at least $HA\land B|B)=0.5$ bits/symbol. But what is the most efficient way to satisfy both users' demands simultaneously? Surprisingly, $0.5$ bits/symbol is also sufficient to simultaneously satisfy the demands of both users. This is accomplished as follows. Let us first consider block length $L=1$ and let $S_{AB}$ represent the value of the broadcast symbol $S$ as a function of the values of $A$ and $B$. Now, when $A=0$, then $B=0$ and $B=1$ produce different values of $A\lor B$. In order for User $1$ to be able to distinguish between the two possibilities, we must have $S_{00}\neq S_{01}$. Similarly, when $B=1$, then $A=0$ and $A=1$ produce different values of $A\land B$, so that in order to satisfy User $2$'s demand, we must have $S_{01}\neq S_{11}$. Subject to these two constraints, i.e., $S_{00}\neq S_{01}$ and $S_{01}\neq S_{11}$ let us assign values to $S_{AB}$ to minimize the number of bits needed to send $S_{AB}$ to both users using Slepian-Wolf coding, i.e., $\max(H(S_{AB}|A), H(S_{AB}|B))$. The  solution for this toy problem gives us $S_{AB}=1$ if $(A,B)=(0,1)$ and $S_{AB}=0$ otherwise. Note that 
\begin{eqnarray}
H(S_{AB}|A)&=&P(A=0)H(S_{AB}|A=0)+P(A=1)H(S_{AB}|A=1) = 0.5 \mbox{ bits/symbol}
\end{eqnarray}
and similarly, $H(S_{AB}|B)=0.5$ bits/symbol. Remarkably, Slepian-Wolf coding allows us to satisfy both users' demands by sending only $0.5$ bits/symbol. Specifically, we consider larger blocks of length $L\rightarrow\infty$,  randomly bin the $2^{L}$ realizations of $S_{AB}^L$ into $2^{L(0.5+\epsilon)}$ bins, and broadcast only\footnote{Note that directly setting $S=S_{AB}$ and operating over block length $L=1$ is the best solution for $L=1$, i.e., $R^*_1=H(w_1,w_2)/H(S_{AB})=H(w_1,w_2)/(2-\frac{3}{4}\log_2 (3))$. However, this is not capacity-achieving because $C_\cb = H(w_1, w_2)/2 > R^*_1$. The example shows explicitly why the problem formulation in \eqref{eq:dec1}-\eqref{eq:dec2} in multi-letter form (arbitrarily large $L\in\mathbb{N}$) cannot be trivially single-letterized by restricting to the case $L=1$. } the bin index as $S$ which requires $H(S)\leq L(0.5+\epsilon)$ bits. Because of the joint asymptotic equipartition property (AEP), User $1$ finds a unique $S_{AB}^L$ sequence that is jointly typical with its side-information sequence $A^L$ with high probability, while User $2$ finds a unique $S_{AB}^L$ sequence that is jointly typical with its side-information sequence $B^L$ with high probability. Thus, rates arbitrarily close to $0.5$ bits per source symbol are achievable.\footnote{Slepian-Wolf  coding with distributed side-information in general may need $\epsilon$-error. However, in our case, since the encoder knows all messages and side-information symbols, centralized coding allows us to achieve zero-error --- for almost all realizations of $(W_1, W_2, W_1', W_2)$ the Slepian-Wolf code works, and for the remaining $\epsilon$-probable realizations, we simply send out $(W_1, W_2)$, which has negligible impact on expected rate, as $\epsilon$ can be chosen to be arbitrarily small.}   Remarkably, $0.5$ bits per source symbol is also optimal because 
\begin{eqnarray}
H(A\lor B\mid A) &=& P(A=0) H(A\lor B \mid A=0) + P(A=1) H(A\lor B \mid A=1)\\
&=& 0.5 H(B) +  0.5 (0) = 0.5 \mbox{ bits/symbol}
\end{eqnarray}
and similarly $H(A\land B\mid B)= 0.5$ bits /symbol. Thus, at least $0.5$ bits/symbol is needed to satisfy either user alone. Fig. \ref{fig:ANDOR} illustrates this toy example.

\subsubsection{Example 4: A Ternary AND/OR Problem}
In order to emphasize the difficulty of the computation broadcast problem in general, suppose we only slightly modify the example as follows. Suppose now that $A,B\in\{0,1,2\}$ are i.i.d. uniform $3$-ary random variables. As the natural extension of the previous example to $3$-ary symbols, let us now define $A\lor B$ as $0$ if $(A,B)= (0,0)$ and $1$ otherwise. Similarly, define $A\land B$ as $1$ if $(A,B)=(1,1)$ and $0$ otherwise. As before, User $1$ knows $A$ and wants $A\lor B$ while User $2$ knows $B$ and wants $A\land B$. Even though this problem is only slightly modified from the previous example for which the capacity was  characterized, the capacity for this modified case seems to be a challenging open problem. 

\subsection{Two Classes of Computation Broadcast Problems}
There are two main classes of computation broadcast problems that we explore in this work -- linear settings and matching problems. These classes are defined next.
\subsubsection{Class I: Linear Computation Broadcast}
Because  computations are  often linear, it is of particular interest to consider the linear version of the computation broadcast problem, denoted linear computation broadcast, or LCB. For LCB,  the defining restriction is that $W_1, W_1', W_2, W_2'$ are arbitrary linear combinations of a basis set of independent symbols from a finite field. Let the basis symbols be specified through the $m\times 1$ column vector ${\bf X}=(x_1; x_2; \cdots; x_m)$, where $x_i, i \in \{1,2,\cdots,m\}$ are i.i.d. uniform symbols from a finite field ${\mathbb F}_{q}$, $q=p^n$, for a prime $p$ and an integer $n$.  Since all symbols are linear combinations of the basis symbols, they are represented by $m\times 1$ vectors of linear combining coefficients. Each message or side-information is then specified in terms of such vectors,
\begin{eqnarray}
{\bf W}_1 &=& {\bf X}^T {\bf V}_1 \label{eq:w1}\\
{\bf W}_1' &=& {\bf X}^T{\bf V}'_1  \label{eq:w1'}\\ 
{\bf W}_2 &=& {\bf X}^T{\bf V}_2  \\
{\bf W}_2' &=& {\bf X}^T{\bf V}'_2  \label{eq:w2'}
\end{eqnarray}
For example, if ${\bf V}_1$ is comprised of two $m\times 1$ vectors, i.e., ${\bf V}_1= [{\bf V}_1^{(1)}, {\bf V}_1^{(2)}]$, then it means that $W_1$ is comprised of symbols ${\bf X}^T{\bf V}_1^{(1)}, {\bf X}^T {\bf V}_2^{(2)}$, and may be represented as ${\bf W}_1=[{\bf X}^T{\bf V}_1^{(1)}, {\bf X}^T {\bf V}_2^{(2)}]$. Note that the broadcast information $S$ is  \emph{not} constrained to be a linear function of the basis symbols, although as we will prove,  it turns out that linear forms of $S$ are information theoretically optimal (refer to Section \ref{sec:linear}). 

\subsubsection{Class II: Matching Problems}
While we are able to characterize the capacity of linear computation broadcast in this work, the capacity for non-linear settings remains open in general. In order to explore the challenges that arise in non-linear settings, we will focus on a limited class of non-linear computation broadcast problems, that we label as `matching' problems. Here, the dependence between $w_1$ and $w_2$ is in the form of an invertible function (a perfect matching, equivalently a permutation) that depends upon $w_1', w_2'$. The dependence is minimal in the sense that each of $(w_1', w_2', w_1)$  and $(w_1', w_2', w_2)$ are independent and uniformly distributed over $[m_1]\times[m_2]\times [m]$. Mathematically,
\begin{align}
(w_1, w_2, w_1', w_2')&\in [m]\times[m]\times[m_1]\times[m_2],\\
H(w_1',w_2',w_1)&=H(w_1')+H(w_2')+H(w_1)=\log_2 (m_1)+\log_2 (m_2)+\log_2 (m),\\
H(w_1',w_2',w_2)&=H(w_1')+H(w_2')+H(w_2)=\log_2 (m_1)+\log_2(m_2)+\log_2(m),\\
H(w_1\mid w_1',w_2', w_2)&=H(w_2\mid w_1',w_2', w_1)=0.
\end{align}
Note that this setting includes both Example 1 and Example 2 as special cases when the matching is reduced to a linear mapping. We will explore how the \emph{structure} of the matching affects the capacity of computation broadcast. In particular, we will characterize both minimally structured and maximally structured cases that correspond to the extremal values of capacity, while all other settings lie somewhere between these extremal values. 

\section{Results}\label{sec:main}
\subsection{A General Converse}
Our first  result is a general converse bound, stated in the following theorem.

\begin{theorem}\label{thm:converse}
{\normalfont[General Converse]}
For any computation broadcast problem,
\begin{eqnarray}
C_\cb ~\leq~ \overline{C}_\cb ~\leq~ \frac{H(w_1,w_2)}{H(w_1|w_1')+H(w_2|w_2') 
-\min\Big(I(w_1; w_2, w_2'|w_1'), I(w_2; w_1, w_1'|w_2')\Big) }.\label{eq:capbound}
\end{eqnarray}
\end{theorem}

The proof of Theorem \ref{thm:converse} is presented in Section \ref{sec:converse}. In fact, the  bound is  intuitively quite obvious. The key to the bound is that $$R_L^* \leq H(W_1,W_2) / \left[ H(W_1|W_1')+H(W_2|W_2') -\min\Big(I(W_1; W_2, W_2'|W_1'), I(W_2; W_1, W_1'|W_2')\Big) \right]$$ 
which follows from the following two bounds.
\begin{eqnarray}
H(S) &\geq& H(W_1\mid W_1')+H(W_2\mid W_1,W_1',W_2')\label{eq:firstbound}\\
H(S) &\geq& H(W_2\mid W_2')+H(W_1\mid W_2,W_2',W_1')\label{eq:secondbound}
\end{eqnarray}
For the first bound in \eqref{eq:firstbound}, note that User $1$, who already knows $W_1'$,  at least needs another $H(W_1|W_1')$ bits to decode $W_1$, and after everything known to User $1$ is given to User $2$ by a genie, User 2, who now knows $W_1, W_1', W_2'$,  needs another $H(W_2|W_1, W_1', W_2')$ bits to decode $W_2$. So without the genie we cannot need any less. The same intuition can be applied with the users switched for \eqref{eq:secondbound}. In fact, the basic intuition is strong enough that the bound holds even in the relaxed entropic formulation, so we also have 
$$ \overline{C}_\cb \leq H(w_1, w_2) / \left[ H(w_1|w_1')+H(w_2|w_2') -\min\Big(I(w_1; w_2, w_2'|w_1'), I(w_2; w_1, w_1'|w_2')\Big) \right]$$ Finally, as discussed previously, $C_\cb \leq \overline{C}_\cb$ is true by definition since the entropic formulation is a relaxation of the complete (structural) formulation of the computation broadcast problem.

What is surprising about the converse bound is that it turns out to be tight for many settings of interest. In particular, for the linear computation broadcast problem, the converse bound is tight for both the entropic formulation as well as the structured formulation, i.e., it is also achievable. For the class of matching problems, the bound is tight for the entropic formulation, but not necessarily for the complete structured formulation, i.e., it is not achievable in general and the capacity may be strictly smaller once the dependency structure of the problem is fully accounted for. This makes sense because the converse bound is based on only entropic inequalities, in fact it uses only Shannon information inequalities, i.e., sub-modularity properties, so it cannot capture more structural constraints than the entropic formulation. 

\subsection{Capacity of Linear Computation Broadcast}\label{sec:lcb}
Our second  result shows that the bound  in Theorem \ref{thm:converse} is tight for the linear computation broadcast problem for any block length $L$. 
We state this result in the following theorem.
\begin{theorem}\label{thm:linear}
For linear computation broadcast, the capacity is
\begin{eqnarray*}
C_\cb ~=~ \overline{C}_\cb ~=~ \frac{H(w_1,w_2)}{H(w_1|w_1')+H(w_2|w_2') 
-\min\Big(I(w_1; w_2, w_2'|w_1'), I(w_2; w_1, w_1'|w_2')\Big)}.
\end{eqnarray*}
\end{theorem}

The proof of Theorem \ref{thm:linear} is presented in Section \ref{sec:linear}. Since the converse is already available from Theorem \ref{thm:converse}, only a proof of achievability is needed. Intuitively, the achievable scheme is described as follows. First without loss of generality it is assumed that $W_1$ is independent of $W_1'$, and similarly, $W_2$ is independent of $W_2'$, because any dependence can be extracted separately as a sub-message that is already available to the user, and therefore can be eliminated from the user's demand. The core of the achievability argument then is that for linear computation broadcast, the problem can be partitioned into $3$ independent sub-problems, labeled $a,b,c$.  Correspondingly, each message is split into $3$ independent parts: ${\bf W}_i=({\bf W}_{ia},{\bf W}_{ib},{\bf W}_{ic})$, $i\in\{1,2\}$. The $3$ partitions are then solved as separate and independent problems, with corresponding solutions ${\bf S}_a, {\bf S}_b, {\bf S}_c$ that ultimately require a total of $H(S)=H({\bf S}_a)+H({\bf S}_b)+H({\bf S}_c)$ bits. The sub-messages ${\bf W}_{1a}, {\bf W}_{2a}$ are analogous to Example $1$, i.e., ${\bf W}_{1a}$ is a function\footnote{In fact ${\bf W}_{1a}$ may be a linear combination of both ${\bf W}_1', {\bf W}_2'$ (see \eqref{eq:v1primev2prime}), but since ${\bf W}_1'$ is already known to User $1$, there is no loss of generality in restricting ${\bf W}_{1a}$ to be the part that only depends on ${\bf W}_2'$. Similarly, there is no loss of generality in restricting ${\bf W}_{2a}$ to a function of ${\bf W}_1'$.} of ${\bf W}_2'$ while ${\bf W}_{2a}$ is a function of ${\bf W}_1'$, so that it suffices to send $H({\bf S}_a)=\max(H({\bf W}_{1a}),H({\bf W}_{2a}))$ bits as in Example $1$. The partition ${\bf W}_{1b},{\bf W}_{2b}$ is analogous to Example $2$, i.e., it satisfies a dependence relation of the form ${\bf W}_{1b}{\bf M}_{1b}+{\bf W}_{2b}{\bf M}_{2b}+{\bf W}_{1}'{\bf M}_1'+{\bf W}_{2}'{\bf M}_2'={\bf 0}$, where $H({\bf W}_{1b})=H({\bf W}_{2b})$, ${\bf M}_1', {\bf M}_2', {\bf M}_{1b}, {\bf M}_{2b}$ are linear transformations (matrices) and ${\bf M}_{1b}, {\bf M}_{2b}$ are invertible. This is solved by sending, ${\bf S}_b={\bf W}_{2b}{\bf M}_{2b}+{\bf W}_2'{\bf M}_2'$ which satisfies the demands of both users and requires $H({\bf S}_b)=H({\bf W}_{1b})= H({\bf W}_{2b})$ bits. Finally, the partition ${\bf W}_{1c}, {\bf W}_{2c}$ is trivial as it is comprised of sub-messages that are independent of each other and of all side-information, so the optimal solution for this part is  simply uncoded transmission ${\bf S}_c=({\bf W}_{1c}, {\bf W}_{2c})$ which takes $H({\bf S}_c)=H({\bf W}_{1c})+H({\bf W}_{2c})$ bits. Without loss of generality, suppose $H({\bf W}_{1a})\geq H({\bf W}_{2a})$. Then, the total number of bits needed is $H(S)=H({\bf S}_a)+H({\bf S}_b)+H({\bf S}_c)=H({\bf W}_{1a})+H({\bf W}_{1b})+H({\bf W}_{1c})+H({\bf W}_{2c})=H({\bf W}_1\mid {\bf W}_1')+H({\bf W}_2\mid {\bf W}_1, {\bf W}_1', {\bf W}_2')$ which matches the converse bound.  Therefore $
H({\bf W}_1, {\bf W}_2)/C_\cb = H({\bf W}_1\mid {\bf W}_1')+H({\bf W}_2\mid {\bf W}_1, {\bf W}_1', {\bf W}_2'))$ in this case. Note that if we assumed instead that $H({\bf W}_{1a})\leq H({\bf W}_{2a})$ then the  number of bits required by the achievable scheme, and the tight converse bound on $H({\bf W}_1, {\bf W}_2)/C_\cb$  
(because it is achievable), would both be equal to 
$H({\bf W}_2\mid {\bf W}_2')+H({\bf W}_1\mid {\bf W}_2, {\bf W}_1',{\bf W}_2')$.

\subsection*{Example}\label{sec:ex}
Let ${\bf X}=[x_1,x_2,x_3,x_4,x_5,x_6,x_7]^T$, whose elements are i.i.d. uniform random variables in $\mathbb{F}_3$. Let us define
\begin{align}
{\bf W}_1' &= [x_1, x_3],&{\bf W}_1 &= [(x_1 + 2x_2),~ (x_3 + x_5),~ (x_1 + x_4 + x_6), ~x_7]\\
{\bf W}_2' &= [x_2, x_4],& {\bf W}_2 &= [(2x_1 + x_2), ~x_5, ~(x_2 + x_4 + 2x_6)] 
\end{align}
Splitting into $a,b,c$ sub-problems (see Section \ref{sec:linear}), we have
\begin{align}
& {\bf W}_{1a} = [x_1+2x_2]\equiv [2x_2], && {\bf W}_{1b} = [x_3+x_5, x_1+x_4+x_6],&& {\bf W}_{1c} = [x_7] \\
& {\bf W}_{2a} = [2x_1+x_2]\equiv [2x_1],&& {\bf W}_{2b} = [x_5, x_2+x_4+2x_6], &&{\bf W}_{2c} = [~]
\end{align}
Following the procedure in Section \ref{sec:linear} we will find that ${\bf W}_{1a}=[x_1+2x_2]$, which makes ${\bf W}_{1a}$ a function of $({\bf W}_1', {\bf W}_2')$. However, note that setting ${\bf W}_{1a}=[x_1+2x_2]$ is equivalent (`$\equiv$')  to setting ${\bf W}_{1a}=[2x_2]$ because User $1$ already knows $x_1$. In the same sense, setting ${\bf W}_{2a}=[2x_1+x_2]$ is equivalent to setting ${\bf W}_{2a}=[2x_1]$ because $x_2$ is already known to User $2$ as side-information. Thus, without loss of generality, ${\bf W}_{1a}$ is a function of only ${\bf W}_2'$, and ${\bf W}_{2a}$ is a function of only ${\bf W}_1'$. Thus, sub-problem `$a$' is analogous to the setting of Example $1$, and is solved by transmitting ${\bf S}_{a}=[2x_2+2x_1]$. For sub-problem `$b$', note that
\begin{align}
& {\bf W}_{1b} 
\left[ \begin{array}{cc}
1 & 0 \\
0 & 2
\end{array}
\right] 
+ {\bf W}_{2b}
\left[ \begin{array}{cc}
-1 & 0 \\
0 & -1
\end{array}
\right] 
+ {\bf W}'_{1} 
\left[ \begin{array}{cc}
0 & -2 \\
-1 & 0
\end{array}
\right] 
+ {\bf W}'_{2} 
\left[ \begin{array}{cc}
0 & 1 \\
0 & -1
\end{array}
\right] = {\bf 0}
\end{align}
and the matrices multiplying ${\bf W}_{1b}$ and ${\bf W}_{2b}$ are invertible matrices. This problem is  analogous to Example $2$ and is solved by sending ${\bf S}_b={\bf W}_{2b}{\bf M}_{2b}+{\bf W}_2'{\bf M}_2'=[-x_5, -2x_4-2x_6]$. Finally, sub-problem `$c$' is trivially solved by sending ${\bf S}_{c}=[{\bf W}_{1c}, {\bf W}_{2c}]=[x_7]$.
Combining ${\bf S}_{a}, {\bf S}_b, {\bf S}_c$ into $S$, we have the solution,
\begin{align}
S& = ((2x_2 + 2x_1), (-x_5), (-2x_4-2x_6), (x_7)) \label{eq:sb1}
\end{align}
which needs $H(S)=4$ symbols from $\mathbb{F}_3$  per block, and the rate achieved is $R = H({\bf W}_1, {\bf W}_2)/ H(S) = 7/4$. Since this matches the converse bound from Theorem \ref{thm:converse}, we have shown that for this example,
\begin{align}
C_\cb = 7/4.
\end{align}

\subsection{Extra-entropic Structure Matters}
Theorem \ref{thm:linear} shows that the general converse of Theorem \ref{thm:converse} is tight for linear computation broadcast, and the solution of the structural formulation in Section \ref{sec:structure} coincides with the solution to the entropic formulation in Section \ref{sec:entropic}, i.e., 
$C_\cb = \overline{C}_\cb$. Our next result shows that this is not the case in general.

\begin{theorem}\label{thm:structure}
There exist instances of the computation broadcast problem where $C_\cb < \overline{C}_\cb$. Thus,  the  converse in Theorem \ref{thm:converse} is not always tight  for the general (non-linear) computation broadcast problem, and extra-entropic structure matters.
\end{theorem}
{\it Proof:}
To prove this, we will present two instances of computation broadcast, say $\mbox{CB}_1, \mbox{CB}_2$, that have the same entropic formulations, so they have the same $\overline{C}_\cb$.  Yet, these two instances have  different structural formulations that produce different capacities. Incidentally, both instances are matching problems.\\
\noindent{\bf CB}$_1$: This instance of the computation broadcast problem is defined by $(w_1', w_2', w_1, w_2) \in \{0,1\}\times\{0,1\}\times\{0,1,2,3\}\times\{0,1,2,3\}$. The marginal distribution of each random variable is uniform over its own alphabet set. Furthermore, $w_1', w_2', w_1$ are independent and $w_2$  is uniquely determined by $w_1', w_2', w_1$ according to the functional relationship,
\begin{eqnarray}
w_2 = ( w_1 + z) \mod 4,
\end{eqnarray}
where  $z$ is a function of $(w_1',w_2')$, defined as follows.
\begin{eqnarray}
\begin{array}{c|c|c|}
z&w_2'=0&w_2'=1\\ \hline
w_1'=0& 0 & 1\\ \hline
w_1'=1& {\color{black}2} & {\color{black}3}\\ \hline
\end{array} \label{eq:ch2}
\end{eqnarray}
Thus, for all $w_1', w_2' \in \{0,1\}, w_1, w_2 \in \{0,1,2,3\}$
\begin{eqnarray}
P_{w_1, w_1', w_2, w_2'} &=& P_{w_1'} P_{w_2'} P_{w_1} P_{w_2 | w_1', w_2', w_1} \\
&=& 1/2 \times 1/2 \times 1/4 \times \mathbbm{1} \Big(w_2 = (w_1 + z) \mod 4\Big) \label{eq:dist}
\end{eqnarray}
where $\mathbbm{1}(x)$ is the indicator function that takes value 1 if the event $x$ is true and 0 otherwise. Note that given $(w_1', w_2')$, there is an invertible mapping between $w_1$ and $w_2$, which makes this a matching problem. The entropies of all subsets of $\{w_1', w_2', w_1, w_2\}$ are found as follows.
\begin{align}
H(w_1')&=H(w_2')=1, ~~
H(w_1)=H(w_2)=2 \label{eq:h1}\\
H(u,v)&=H(u)+H(v), ~\forall \{u,v\}\subset\{w_1', w_2', w_1, w_2\}\\
H(t,u,v)&=4, ~\forall \{t,u,v\}\subset\{w_1', w_2', w_1, w_2\}\\
H(w_1, w_1', w_2, w_2') &= 4 \label{eq:h4}
\end{align}
Theorem \ref{thm:converse} establishes a converse bound for this problem,
$C_\cb \leq \overline{C}_\cb \leq 2$. The bound turns out to be achievable by setting $L=1$ and choosing $S = (w_1 + 2w_1') \mod 4$ which satisfies both users' demands. This is verified as follows. User $1$ obtains $w_1$ by computing $w_1=(S-2w_1')\mod 4$.  User $2$ obtains $w_2$ by computing $w_2=(S+w_2')\mod 4$, which is possible because in this problem $z=(2w_1'+w_2')\mod 4$. Since $H(S)=2$ bits  and the rate achieved is $H(w_1,w_2)/H(S) = 2$, the achievability matches the converse, which proves that for CB$_1$,  
the capacity $C_{\cb_1}=2$.

\noindent{{\bf CB}$_2$:} CB$_2$ is identical to CB$_1$ in all respects, except that the definition of $z$ is slightly modified as follows.
\begin{eqnarray}
\begin{array}{c|c|c|}
z&w_2'=0&w_2'=1\\ \hline
w_1'=0& 0 & 1\\ \hline
w_1'=3& {\color{black}3} & {\color{black}2}\\ \hline
\end{array} \label{eq:ch1}
\end{eqnarray}
The change in the $z$ does not affect the entropic formulation of the problem. It is easily verified that the entropies of all subsets of $\{w_1',w_2',w_1,w_2\}$ are still given by \eqref{eq:h1}-\eqref{eq:h4}. Since the entropic formulation is not affected we must still  $\overline{C}_{\cb_1} =\overline{C}_{\cb_2} = 2$.
However, the following lemma claims that 
the capacity  $C_{\cb_2}=\frac{4}{4-\log_2(3)}$ is strictly smaller than $C_{\cb_1}$, i.e., Theorem \ref{thm:structure} is proved and the extra-entropic structure reduces capacity in this case.
\begin{lemma}\label{lemma:extra}
For the computation broadcast problem CB$_2$ defined above,
\begin{align}
C_{\cb_2}&=\frac{4}{4-\log_2(3)}
\end{align} 
\end{lemma}
The proof of Lemma \ref{lemma:extra} is presented in Section \ref{sec:exproof}.\hfill\QED

\subsection{Capacity of Matching Computation Broadcast}
To gain a deeper understanding of the significance of extra-entropic structure that is revealed by CB$_1$ and CB$_2$, we explore the capacity of a class of computation broadcast problems called matching problems, which include CB$_1$ and CB$_2$ as special cases. For matching problems we have $(w_1,w_2, w_1',w_2')\in[m_1]\times[m_2]\times[m]\times[m]$ where $m_1, m_2, m\in\mathbb{N}$. The tuple $(w_1', w_2', w_1)$ is uniformly distributed over $[m_1]\times[m_2]\times[m]$, while $w_2$ is a function of $w_1', w_2', w_1$ defined as,
\begin{eqnarray}
w_2&=&\pi_{w_1', w_2'}(w_1) \label{eq:function}
\end{eqnarray}
where $\pi_{w_1',w_2'}$ is a permutation on $[m]$ that depends on the realization of the side-information $(w_1', w_2')$. Distinct realizations of $(w_1',w_2')$ may or may not produce distinct permutations.
$\pi_{w_1',w_2'}$ may be represented in a matrix form as follows.
\begin{eqnarray*}
\begin{array}{c|c|c|c|c|}
\pi_{w_1', w_2'} &w_2'=1&w_2'=2&\cdots&w_2'=m_2\\ \hline
w_1'=1& \pi_{1,1} & \pi_{1,2}&\cdots&\pi_{1,m_2}\\ \hline
w_1'=2& \pi_{2,1}  &\pi_{2,2} &\cdots& \pi_{2,m_2} \\ \hline
\vdots& \vdots &\vdots&\cdots&\vdots \\ \hline
w_1'=m_1&\pi_{m_1, 1}  &\pi_{m_1,2}&\cdots&\pi_{m_1,m_2} \\ \hline
\end{array}
\end{eqnarray*}
Let this matrix be denoted by $\Pi$. Specification of $\Pi$ completely defines the structure of the matching computation broadcast problem.
For all $w_1' \in [m_1], w_2' \in [m_2], w_1, w_2 \in [m]$, we have
\begin{eqnarray}
P_{w_1, w_1', w_2, w_2'} &=& P_{w_1'} P_{w_2'} P_{w_1} P_{w_2 | w_1', w_2', w_1} \\
&=& 1/m_1 \times 1/m_2 \times 1/m \times \mathbbm{1} (w_2 = \pi_{w_1',w_2'}(w_1)) \label{eq:per}
\end{eqnarray}
Note that $w_1', w_2', w_2$ are independent. 

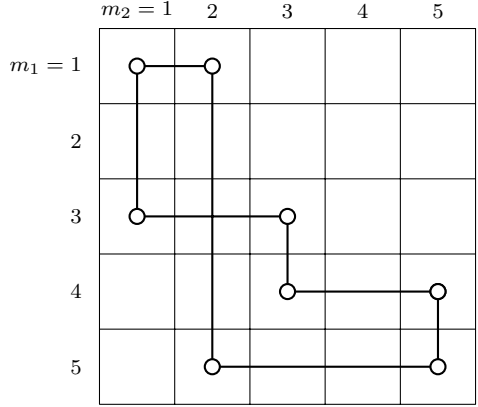
\begin{figure}
\begin{center}
\begin{tikzpicture}
\draw [black,step=1cm] grid (5,5);
\draw [black, thick] 
(0.5,4.5) 
- - (0.5, 2.5) 
- -  (2.5, 2.5) 
- - (2.5, 1.5)
- - (4.5, 1.5)
- - (4.5, 0.5)
- - (1.5, 0.5)
- - (1.5, 4.5)
- - cycle;
\draw[black, thick, fill=white] (0.5,4.5) circle (0.1cm);
\draw[black, thick, fill=white] (1.5,4.5) circle (0.1cm);
\draw[black, thick, fill=white] (1.5,0.5) circle (0.1cm);
\draw[black, thick, fill=white] (4.5,0.5) circle (0.1cm);
\draw[black, thick, fill=white] (4.5,1.5) circle (0.1cm);
\draw[black, thick, fill=white] (4.5,1.5) circle (0.1cm);
\draw[black, thick, fill=white] (2.5,1.5) circle (0.1cm);
\draw[black, thick, fill=white] (2.5,2.5) circle (0.1cm);
\draw[black, thick, fill=white] (0.5,2.5) circle (0.1cm);
\node  [left] at (-0.1,4.5) {\scriptsize $m_1=1$};
\node  [left] at (-0.1,3.5) {\scriptsize $2$};
\node  [left] at (-0.1,2.5) {\scriptsize $3$};
\node  [left] at (-0.1,1.5) {\scriptsize $4$};
\node  [left] at (-0.1,0.5) {\scriptsize $5$};
\node  [above] at (0.5,5) {\scriptsize $m_2=1$};
\node  [above] at (1.5,5) {\scriptsize $2$};
\node  [above] at (2.5,5) {\scriptsize $3$};
\node  [above] at (3.5,5) {\scriptsize $4$};
\node  [above] at (4.5,5) {\scriptsize $5$};

 \end{tikzpicture}
\end{center}
\caption{\small A cycle, $(1,1)\leftrightarrow(1,2)\leftrightarrow (5,2)\leftrightarrow (5,5) \leftrightarrow (4,5) \leftrightarrow (4,3)\leftrightarrow (3,3)\leftrightarrow (3,1)\leftrightarrow(1,1)$.}
\label{fig:cycle}
\end{figure}

Next let us introduce some definitions that are useful to gauge the amount of structure in a given $\Pi$. We begin with the notion of a cycle, which is a closed path on an $m_1\times m_2$ grid, obtained by a sequence of alternating horizontal and vertical steps. See Fig. \ref{fig:cycle} for an illustration. 
\begin{definition}[Cycle] Let $N\geq 4$ be an even number. We say that the $N$ terms, $(a_1, b_1)$, $(a_2, b_2)$, $\cdots$, $(a_N, b_N)\in[m_1]\times[m_2]$,  form a cycle of length $N$ in $[m_1]\times[m_2]$, denoted by 
\begin{eqnarray}
(a_1,b_1) \leftrightarrow (a_2,b_2)\leftrightarrow\cdots\leftrightarrow (a_N,b_N)\leftrightarrow (a_1,b_1)
\end{eqnarray}
if both of the following properties are true $\forall i\in [N]$:
\begin{enumerate}
\item $a_i=a_{i+1}$ and $b_i\neq b_{i+1}$ if $i$ is odd.
\item $b_i=b_{i+1}$ and $a_i\neq a_{i+1}$ if $i$ is even.
\end{enumerate}
where we interpret all indices modulo $N$ (so, e.g., $a_{N+1}=a_1$).
\end{definition}
Other descriptions are also possible for the same cycle. For example, the cycle in Fig. \ref{fig:cycle} can also be identified as $(5,2)\leftrightarrow (5,5) \leftrightarrow (4,5) \leftrightarrow (4,3)\leftrightarrow (3,3)\leftrightarrow (3,1)\leftrightarrow(1,1)\leftrightarrow(1,2)\leftrightarrow(5,2)$.

\begin{definition}[Induced Permutation] For a cycle $(a_i, b_i)_{i\in[N]}$, we define its induced permutation as 
\begin{align}
\pi_{a_1,b_1}\pi_{a_2,b_2}^{-1}\pi_{a_3,b_3}\pi_{a_4,b_4}^{-1}\cdots \pi_{a_N,b_N}^{-1}
\end{align}
\end{definition}

\begin{definition}[Maximally Structured] We say that $\Pi$ is maximally structured if  the induced permutation for every possible cycle in $[m_1]\times[m_2]$ is the identity.\footnote{A permutation $\pi$ on $[m]$ is the identity if and only if it maps every element to itself, i.e., $\pi[i]=i$ for all $i\in[m]$.}
\end{definition}

\begin{definition}[Minimally Structured] We say that $\Pi$ is minimally structured if the induced permutation for every possible cycle in $[m_1]\times[m_2]$ is a derangement.\footnote{A permutation $\pi$ on $[m]$ is a derangement if and only if no element is mapped to itself, i.e., $\pi[i]\neq i$ for all $i\in[m]$.
}
\end{definition}

Maximal structure is a generalization of the setting in CB$_1$. In CB$_1$ there is only one possible cycle: $(1,1) \leftrightarrow (1,2) \leftrightarrow (2,2) \leftrightarrow (2,1) \leftrightarrow (1,1)$, for which the induced permutation $\pi_{1,1} \pi_{1,2}^{-1} \pi_{2,2} \pi_{2,1}^{-1}$ is the identity. 
Minimal structure is a generalization of the setting in CB$_2$. For the cycle $(1,1) \leftrightarrow (1,2) \leftrightarrow (2,2) \leftrightarrow (2,1) \leftrightarrow (1,1)$, the induced permutation $\pi_{1,1} \pi_{1,2}^{-1} \pi_{2,2} \pi_{2,1}^{-1}$ is a derangement. 

The significance of this structure is revealed by the next theorem.

\begin{theorem} \label{thm:str}
For a matching computation broadcast problem specified by the structure $\Pi$, 
\begin{eqnarray*}
\frac{2\log_2(m)}{\log_2(m)+\log_2(m_1m_2)-\log_2(m_1+m_2-1)} \leq  C_\cb \leq  2.
\end{eqnarray*}
The upper bound is tight {\color{black}if} $\Pi$ is maximally structured. The lower bound is tight if $\Pi$ is minimally structured.
\end{theorem}

The proof of Theorem \ref{thm:str} is presented in Section \ref{sec:str}. 
The following observations are in order.
\begin{enumerate}
\item Since maximally structured settings represent the best case and minimally structured settings the worst case, it is evident that structure is beneficial. 
\item  The proof presented in Section \ref{sec:str} shows that  the minimally structured setting still has some (unavoidable) combinatoric structure that is critical for the optimal achievable scheme.
\item To contrast with the previous observation, consider the following. Suppose $m_1=m_2\triangleq m'$ and all alphabet sizes grow together proportionately. Then the minimally structured setting essentially loses all its structure and random binning is close to optimal. To see this, consider the term $\log_2(m_1m_2)-\log_2(m_1+m_2-1)$. For large values of $m'$, this becomes $\approx 2\log_2(m')-\log_2(2m')=\log_2(m')-1=H(w_1')-1$. So the capacity $C_{\cb}$ approaches the value $H(w_1,w_2)/[H(w_1)+H(w_1')]$ which is achievable\footnote{It is achieved by separately compressing and sending $w_1, w_1'$. User $1$ directly receives $w_1$ and User $2$ decodes $w_2=\pi_{w_1',w_2'}(w_1)$. } by random binning. Thus, random binning is asymptotically optimal for minimally structured instances of matching computation broadcast.
\end{enumerate}

\section{Proof of Theorem \ref{thm:converse}: A General Converse}\label{sec:converse}
The converse in Theorem \ref{thm:converse} consists of the following two bounds.
\begin{eqnarray}
H(S) &\geq& H(W_1|W_1') + H(W_2|W_2')  - I(W_2; W_1, W_1'|W_2') \\
&=&H(W_1|W_1') + H(W_2|W_1,W_1',W_2')  \label{eq:c1}\\
H(S) &\geq& H(W_2|W_2')+H(W_1|W_2,W_2',W_1') \label{eq:c2}
\end{eqnarray}
We only need to prove (\ref{eq:c1}), as the proof of (\ref{eq:c2}) follows from symmetry. The proof of (\ref{eq:c1}) is presented next. Note that in the proofs, the relevant equations needed to justify each step are specified by the equation numbers set on top of the (in)equality symbols.

We expand the joint entropy  $H(S,W_1|W_1')$ in two different ways. On the one hand, we have
\begin{eqnarray}
H(S,W_1|W_1')&=&H(S|W_1')+H(W_1|W_1', S) \\
&\overset{(\ref{eq:dec1})}{\leq}& H(S) \label{eq:s1}
\end{eqnarray}
On the other hand, we have
\begin{eqnarray}
&&H(S,W_1|W_1')\\
&=&H(W_1|W_1')+H(S|W_1, W_1')\\
&\geq&H(W_1|W_1')+H(S|W_1, W_1')-H(S|W_1,W_1', W_2, W_2')\\
&=&H(W_1|W_1')+I(S; W_2, W_2'|W_1, W_1')\\
&=&H(W_1|W_1')+H(W_2, W_2'|W_1, W_1')-H(W_2, W_2'|W_1, W_1', S)\\
&=& H(W_1|W_1')+H(W_2'|W_1, W_1')+H(W_2|W_1, W_1',W_2') - H(W_2'|W_1, W_1',S) \notag\\
&&~-H(W_2|W_1, W_1', W_2',S)\\
&\overset{(\ref{eq:dec2})}{=}&H(W_1|W_1')+H(W_2|W_1,W_1',W_2')+ I(S;W_2'|W_1,W_1')\\
&\geq&H(W_1|W_1')+H(W_2|W_1,W_1',W_2') \label{eq:s2}
\end{eqnarray}
Thus combining (\ref{eq:s1}) and (\ref{eq:s2}), we have the desired bound (\ref{eq:c1}). The proof of Theorem \ref{thm:converse} is complete.

\section{Proof of Theorem \ref{thm:linear}: Linear Achievability}\label{sec:linear}
Without loss of generality we will assume that $W_1$ is independent of $W_1'$, and similarly, $W_2$ is independent of $W_2'$. There is no loss of generality in this assumption because any linear dependence between $W_1$ and $W_1'$, or between $W_2$ and $W_2'$, can be extracted separately as a sub-message that is already available to the user, and therefore can be eliminated from the user's demand.

Recall that ${\bf X}=(x_1; x_2; \ldots; x_m)$ is an $m\times 1$ random vector, whose elements $x_i$ are i.i.d. uniform over $\mathbb{F}_q$. All entropies in this section are measured in units of $q$-ary symbols.
 For any matrix ${\bf A}\in\mathbb{F}_q^{m\times n}$, we will use the notation $A$ to denote the set of  column vectors of ${\bf A}$.

\begin{lemma} \label{lemma:obvious} For an arbitrary $m\times n$ matrix ${\bf A}\in\mathbb{F}_q^{m\times n}$, $H({\bf X}^T{\bf A})=\mbox{rank}({\bf A}).$
\end{lemma}

\begin{definition}[Independent subspaces]
Subspaces $\mathcal{A},\mathcal{B}\subset \mathbb{F}_q^m$ are independent if $\mathcal{A}\cap \mathcal{B}=\{\bf 0\}$.
\end{definition}

\begin{lemma}\label{lemma:ind}
For arbitrary matrices ${\bf A}\in\mathbb{F}_q^{m\times n_A}, {\bf B}\in\mathbb{F}_q^{m\times n_B}$, the mutual information $I({\bf X}^T{\bf A}; {\bf X}^T{\bf B})=0$
if and only if $\mbox{span}(A)$ and $\mbox{span}(B)$ are independent subspaces.
\end{lemma}
The proofs of Lemma \ref{lemma:obvious} and Lemma \ref{lemma:ind} are immediate and are deferred to the Appendix.

\noindent Define
\begin{align}
{\bf V}_{1a}&=\left[{\bf V}_{1a}^{(1)},{\bf V}_{1a}^{(2)},\cdots,{\bf V}_{1a}^{(n_{1a})} \right]\in\mathbb{F}_q^{m\times n_{1a}},&&{\bf V}_{2a}&=\left[{\bf V}_{2a}^{(1)},{\bf V}_{2a}^{(2)},\cdots,{\bf V}_{2a}^{(n_{2a})} \right]\in\mathbb{F}_q^{m\times n_{2a}}\\
{\bf V}_{1b}&=\left[{\bf V}_{1b}^{(1)},{\bf V}_{1b}^{(2)},\cdots,{\bf V}_{1b}^{(n_{1b})} \right]\in\mathbb{F}_q^{m\times n_{1b}},&& {\bf V}_{2b}&=\left[{\bf V}_{2b}^{(1)},{\bf V}_{2b}^{(2)},\cdots,{\bf V}_{2b}^{(n_{2b})} \right]\in\mathbb{F}_q^{m\times n_{2b}}\\
{\bf V}_{1c}&=\left[{\bf V}_{1c}^{(1)},{\bf V}_{1c}^{(2)},\cdots,{\bf V}_{1c}^{(n_{1c})} \right]\in\mathbb{F}_q^{m\times n_{1c}},&&{\bf V}_{2c}&=\left[{\bf V}_{2c}^{(1)},{\bf V}_{2c}^{(2)},\cdots,{\bf V}_{2c}^{(n_{2c})} \right]\in\mathbb{F}_q^{m\times n_{2c}}
\end{align}
such that
\begin{enumerate}
\item $V_{1a}, V_{1b}, V_{1c}$ are disjoint sets.
\item $V_{1a}$ is a basis for $\mbox{span}(V_1)\cap\mbox{span}(V_1'\cup V_2')$.
\item $V_{1a}\cup V_{1b}$ is a basis for $\mbox{span}(V_1)\cap\mbox{span}(V_1'\cup V_2'\cup V_2)$.
\item $V_{1a}\cup V_{1b}\cup V_{1c}$ is a basis for $\mbox{span}(V_1)$.
\item $V_{2a}, V_{2b}, V_{2c}$ are disjoint sets.
\item $V_{2a}$ is a basis for $\mbox{span}(V_2)\cap\mbox{span}(V_1'\cup V_2')$.
\item $V_{2a}\cup V_{2b}$ is a basis for $\mbox{span}(V_2)\cap\mbox{span}(V_1'\cup V_2'\cup V_1)$.
\item $V_{2a}\cup V_{2b}\cup V_{2c}$ is a basis for $\mbox{span}(V_2)$.
\end{enumerate}
Recall that basis vectors must be linearly independent. The existence of such $V_{ia}, V_{ib}, V_{ic}$, $i\in\{1,2\}$, follows from Steinitz exchange lemma which guarantees that given a set of basis vectors $\{{\bf p}_1, {\bf p}_2, \cdots, {\bf p}_k\}$ for a $k$-dimensional subspace $\mathcal{P}$, and an arbitrary $m$-dimensional vector space $\mathcal{Q}$, such that $\mathcal{P}\subset \mathcal{Q}$, there exist ${\bf q}_1,\cdots, {\bf q}_{m-k}\in \mathcal{Q}\backslash \mathcal{P}$ such that  $\{{\bf p}_1, {\bf p}_2, \cdots, {\bf p}_k, {\bf q}_1, {\bf q}_2, \cdots, {\bf q}_{m-k}\}$ is a basis for $\mathcal{Q}$.

{\it Remark: As an illustration of this construction, consider the  example presented  in Section \ref{sec:lcb} where we have,
\begin{align}
&{\bf X}=[x_1,x_2,x_3,x_4,x_5,x_6,x_7]^T\\
&{\bf W}_1' = [x_1, x_3], ~{\bf W}_1 = [x_1 + 2x_2, x_3 + x_5, x_1 + x_4 + x_6, x_7]\\
&{\bf W}_2' = [x_2, x_4], ~{\bf W}_2 = [2x_1 + x_2, x_5, x_2 + x_4 + 2x_6] 
\end{align}
This gives us,
\begin{align}
{\bf V}_1'&=[{\bf e}_1, {\bf e}_3],& {\bf V}_1&=[{\bf e}_1+2{\bf e}_2, {\bf e}_3+{\bf e}_5, {\bf e}_1+{\bf e}_4+{\bf e}_6, {\bf e}_7] \label{eq:vv1}\\
{\bf V}_2'&=[{\bf e}_2, {\bf e}_4],& {\bf V}_2&=[2{\bf e}_1+{\bf e}_2, {\bf e}_5,{\bf e}_2+{\bf e}_4+2{\bf e}_6]
\end{align}
and
\begin{align}
{\bf V}_{1a}&=[{\bf e}_1+2{\bf e}_2],&{\bf V}_{2a}&=[2{\bf e}_1+{\bf e}_2] \\
{\bf V}_{1b}&=[{\bf e}_3+{\bf e}_5, {\bf e}_1+{\bf e}_4+{\bf e}_6],& {\bf V}_{2b}&=[{\bf e}_5,{\bf e}_2+{\bf e}_4+2{\bf e}_6] \\
{\bf V}_{1c}&=[{\bf e}_7],&{\bf V}_{2c}&=[~]\label{eq:vv2}
\end{align}
where ${\bf e}_i$ denotes the $i^{th}$ column of the $7\times 7$ identity matrix. }

Next, for $i\in\{1,2\}$ and $\{i,i^c\}=\{1,2\}$, define ${\bf W}_{ia}={\bf X}^T{\bf V}_{ia}$, ${\bf W}_{ib}={\bf X}^T{\bf V}_{ib}$, ${\bf W}_{ic}={\bf X}^T{\bf V}_{ic}$,   so that
\begin{align}
H({\bf W}_{ia}, {\bf W}_{ib}, {\bf W}_{ic})&=H({\bf W}_i)\label{eq:one}\\
H({\bf W}_{ia})+H({\bf W}_{ib})+H({\bf W}_{ic})&=H({\bf W}_i)\label{eq:two}\\
n_{ia}+n_{ib}+n_{ic}&=H({\bf W}_i)\label{eq:three}\\
H({\bf W}_{ia})&=n_{ia}\label{eq:four}\\
H({\bf W}_{ia}\mid {\bf W}_1', {\bf W}_2')&=0\label{eq:five}\\
H({\bf W}_{ib}\mid {\bf W}_1', {\bf W}_2')&=n_{ib}\label{eq:six}\\
H({\bf W}_{ia}, {\bf W}_{ib}\mid {\bf W}_1', {\bf W}_2', {\bf W}_{i^c})&=0\label{eq:seven}\\
H({\bf W}_{ic}\mid {\bf W}_1', {\bf W}_2', {\bf W}_{i^c})&=n_{ic}\label{eq:eight}\\
H({\bf W}_{ia}, {\bf W}_{ib}, {\bf W}_{ic}\mid {\bf W}_1', {\bf W}_2',{\bf W}_1, {\bf W}_2)&=0\label{eq:nine}
\end{align}
\eqref{eq:one} follows from the fact that $V_{ia}\cup V_{ib}\cup V_{ic}$ is the basis for the space spanned by $V_i$, which makes ${\bf W}_i$ an invertible function of $({\bf W}_{ia}, {\bf W}_{ib}, {\bf W}_{ic})$. \eqref{eq:two}-\eqref{eq:four} follow from  Lemma \ref{lemma:obvious} and the fact that the $n_{ia}+n_{ib}+n_{ic}$ vectors in $V_{ia}\cup V_{ib}\cup V_{ic}$ form a basis, so they are linearly independent. 
\eqref{eq:five} holds because $V_{ia}\subset\mbox{span}(V_1'\cup V_2')$, which makes ${\bf W}_{ia}$ a function of ${\bf W}_1', {\bf W}_2'$. \eqref{eq:six} holds because $\mbox{span}(V_{ib})$  is independent of $\mbox{span}(V_1'\cup V_2')$. This is because if a non-zero vector ${\bf U}\in\mbox{span}(V_{ib})\cap\mbox{span}(V_1'\cup V_2')$  then ${\bf U}\in \mbox{span}(V_{ib})$ and ${\bf U}\in\mbox{span}(V_{ia})$, i.e.,  $V_{ib}$ and $V_{ia}$ do not span independent spaces, so $V_{ia}\cup V_{ib}\cup V_{ic}$ cannot be a set of basis vectors.  Similarly, \eqref{eq:eight} holds because $V_{ic}$ and $V_1'\cup V_2'\cup V_{i^c}$ span independent spaces (otherwise $V_{ic}$ and $V_{ib}$ cannot span independent spaces). \eqref{eq:seven} holds because $V_{ia}\cup V_{ib}\subset\mbox{span}(V_1'\cup V_2'\cup V_{i^c})$, and \eqref{eq:nine}  holds because $V_{ia}\cup V_{ib} \cup V_{ic}\subset \mbox{span}(V_1'\cup V_2'\cup V_1\cup V_2)$.

Since $V_{1b}\subset \mbox{span}(V_1'\cup V_2'\cup V_2)$,  there exist matrices ${\bf M}_1'\in\mathbb{F}_q^{n_1'\times n_{1b}}$, ${\bf M}_2'\in\mathbb{F}_q^{n_2'\times n_{1b}}$, ${\bf M}_{2a}\in\mathbb{F}_q^{n_{2a}\times n_{1b}}$, ${\bf M}_{2b}\in\mathbb{F}_q^{n_{2b}\times n_{1b}}$, ${\bf M}_{2c}\in\mathbb{F}_q^{n_{2c}\times n_{1b}}$, such that
\begin{align}
{\bf V}_{1b}&={\bf V}_1'{\bf M}_1'+{\bf V}_2'{\bf M}_2'+{\bf V}_{2a}{\bf M}_{2a}+{\bf V}_{2b}{\bf M}_{2b}+{\bf V}_{2c}{\bf M}_{2c}.\label{eq:key}
\end{align}
We will now show that without loss of generality,  ${\bf M}_{2a}, {\bf M}_{2c}$ are zero matrices, and ${\bf M}_{2b}$ is an invertible square matrix. Since ${\bf V}_{2a}$ can be expanded as a linear combination of ${\bf V}_1'$ and ${\bf V}_2'$, and absorbed into corresponding terms in \eqref{eq:key}, there is no loss of generality in the assumption that ${\bf M}_{2a}$ is the zero matrix, i.e., a matrix  whose elements are all zeros. Next, without loss of generality, we can also assume ${\bf M}_{2c}$ is a zero matrix because $\mbox{span}(V_{2c})$ and $\mbox{span}(V_1'\cup V_2'\cup V_{2b}\cup V_{1b})$ are independent subspaces. This is because of Lemma \ref{lemma:ind} and the fact that ${\bf W}_{2c}$ is independent of $({\bf W}_1', {\bf W}_2', {\bf W}_{2b}, {\bf W}_{1b})$ as shown below.
\begin{align}
I({\bf W}_{2c}; {\bf W}_1', {\bf W}_2', {\bf W}_{2b}, {\bf W}_{1b})&=H({\bf W}_{2c})-H({\bf W}_{2c}\mid {\bf W}_1', {\bf W}_2', {\bf W}_{2b}, {\bf W}_{1b})\\
&=n_{2c}-H({\bf W}_{2c}\mid {\bf W}_1', {\bf W}_2', {\bf W}_{2b}, {\bf W}_{1b})\\
&\leq n_{2c}-H({\bf W}_{2c}\mid {\bf W}_1', {\bf W}_2', {\bf W}_{2b}, {\bf W}_{1})\label{eq:aone}\\
&= n_{2c}-H({\bf W}_{2c}\mid {\bf W}_1', {\bf W}_2',  {\bf W}_{1})\label{eq:atwo}\\
&=0.
\end{align}
\eqref{eq:aone} holds because $V_{1b}\subset \mbox{span}(V_1)$, which makes ${\bf W}_{1b}$  a function of ${\bf W}_1$, while \eqref{eq:atwo} holds because  according to \eqref{eq:seven} ${\bf W}_{2b}$ is a function of ${\bf W}_1', {\bf W}_2', {\bf W}_1$. Thus, without loss of generality \eqref{eq:key} reduces  to
\begin{align}
{\bf V}_{1b}&={\bf V}_1'{\bf M}_1'+{\bf V}_2'{\bf M}_2'+{\bf V}_{2b}{\bf M}_{2b}.\label{eq:keynew}
\end{align}
Next, let us prove that ${\bf M}_{2b}$ is a square matrix, i.e., $n_{1b}=n_{2b}$.
\begin{align}
I({\bf W}_1; {\bf W}_2\mid {\bf W}_1', {\bf W}_2')&=H({\bf W}_1\mid {\bf W}_1', {\bf W}_2')-H({\bf W}_1\mid {\bf W}_2, {\bf W}_1', {\bf W}_2')\\
&=n_{1b}+n_{1c}-n_{1c}\\
&=n_{1b}
\intertext{and similarly,}
I({\bf W}_1; {\bf W}_2\mid {\bf W}_1', {\bf W}_2')&=H({\bf W}_2\mid {\bf W}_1', {\bf W}_2')-H({\bf W}_2\mid {\bf W}_1, {\bf W}_1', {\bf W}_2')\\
&=n_{2b}+n_{2c}-n_{2c}\\
&=n_{2b}
\end{align}
Therefore, $n_{1b}=n_{2b}\triangleq n_b$ and ${\bf M}_{2b}$ is a square matrix. Next, let us prove that it has full rank. Suppose on the contrary that ${\bf M}_{2b}{\bf U}={\bf 0}$ for some ${\bf U}\in\mathbb{F}_q^{n_b\times 1}$ which is not the zero vector. Then \eqref{eq:key} implies that ${\bf V}_{1b}{\bf U}={\bf V}_1'{\bf M}_1'{\bf U}+{\bf V}_2'{\bf M}_2'{\bf U}\in\mbox{span}(V_{1a})$. But ${\bf V}_{1b}{\bf U}$ also belongs to $\mbox{span}(V_{1b})$. Since $\mbox{span}(V_{1a})$ and $\mbox{span}(V_{1b})$ are independent subspaces, we must have ${\bf V}_{1b}{\bf U}={\bf 0}$. This is a contradiction because ${V}_{1b}$ is comprised of linearly independent vectors (because it is a basis), and ${\bf U}$ is not the zero vector. The contradiction proves that ${\bf M}_{2b}$ must have full rank, i.e., it must be invertible.

\noindent {\it Remark: For the example presented in Section \ref{sec:lcb} and matrices specified in (\ref{eq:vv1}) - (\ref{eq:vv2}), we have
\begin{align}
{\bf V}_{1b}&= [{\bf e}_3+{\bf e}_5, {\bf e}_1+{\bf e}_4+{\bf e}_6]\\
&= [{\bf e}_1, {\bf e}_3] \left[ \begin{array}{cc}
0 & 1 \\
1 & 0
\end{array}
\right]  + [{\bf e}_2, {\bf e}_4] 
\left[ \begin{array}{cc}
0 & -\frac{1}{2} \\
0 & \frac{1}{2}
\end{array}
\right]
+ [{\bf e}_5,{\bf e}_2+{\bf e}_4+2{\bf e}_6] 
\left[ \begin{array}{cc}
1 & 0 \\
0 & \frac{1}{2}
\end{array}
\right]\\
&= {\bf V}_1'{\bf M}_1'+{\bf V}_2'{\bf M}_2'+{\bf V}_{2b}{\bf M}_{2b}
\end{align}
}

Without loss of generality, suppose $n_{1a}\geq n_{2a}$. Since $V_{1a}\subset\mbox{span}(V_1'\cup V_2')$, there exist ${\bf P}_1'\in\mathbb{F}_q^{n_1'\times n_{1a}}$, ${\bf P}_2'\in\mathbb{F}_q^{n_2'\times n_{1a}}$, such that the $m\times n_{1a}$ matrix
\begin{align}
{\bf V}_{1a}&={\bf V}_1'{\bf P}_{1}'+{\bf V}_2'{\bf P}_2'.\label{eq:v1primev2prime}
\end{align}
Similarly, there exist ${\bf Q}_1'\in\mathbb{F}_q^{n_1'\times n_{1a}}$, ${\bf Q}_2'\in\mathbb{F}_q^{n_2'\times n_{1a}}$, such that the $m\times n_{1a}$ matrix
\begin{align}
\left[{\bf V}_{2a}, {\bf 0}_{m\times (n_{1a}-n_{2a})}\right]&={\bf V}_1'{\bf Q}_{1}'+{\bf V}_2'{\bf Q}_2'.
\end{align}
Note that $n_{1a}-n_{2a}$ columns of zeros are appended to ${\bf V}_{2a}$ to create a matrix the same size as ${\bf V}_{1a}$.

The transmitted vector ${\bf S}\in\mathbb{F}_q^{(n_{1a}+n_{1b}+n_{1c}+n_{2c})\times 1}$ is now specified as
\begin{align}
{\bf S}&=\left(\underbrace{{\bf X}^T({\bf V}_1'{\bf Q}_1'+{\bf V}_2'{\bf P}_2')}_{{\bf S}_a: ~1\times n_{1a}}, ~ \underbrace{{\bf X}^T({\bf V}_{2b}{\bf M}_{2b}+{\bf V}_2'{\bf M}_2')}_{{\bf S}_b:~ 1\times n_{1b}}, ~\underbrace{{\bf X}^T{\bf V}_{1c}, {\bf X}^T{\bf V}_{2c}}_{{\bf S}_c:~(1\times n_{1c}), (1\times n_{2c})}  \right)^T
\end{align}

{\it Remark: For the example presented in Section \ref{sec:lcb} and matrices specified in (\ref{eq:vv1}) - (\ref{eq:vv2}), we have
\begin{align}
{\bf V}_{1a}&= [{\bf e}_1 + 2{\bf e}_2] = [{\bf e}_1, {\bf e_3}] 
\left[\begin{array}{cc}
1\\
0 
\end{array}
\right] + [{\bf e}_2, {\bf e}_4] 
\left[\begin{array}{cc}
2\\
0  
\end{array}
\right] = {\bf V}_1'{\bf P}_{1}'+{\bf V}_2'{\bf P}_2' \\
{\bf V}_{2a}&= [2{\bf e}_1 + {\bf e}_2] = [{\bf e}_1, {\bf e_3}] 
\left[\begin{array}{cc}
2\\
0 
\end{array}
\right] + [{\bf e}_2, {\bf e}_4] 
\left[\begin{array}{cc}
1\\
0  
\end{array}
\right] = {\bf V}_1'{\bf Q}_{1}'+{\bf V}_2'{\bf Q}_2' 
\\
{\bf S}_a &= {\bf X}^T ({\bf V}_1'{\bf Q}_1'+{\bf V}_2'{\bf P}_2') = {\bf X}^T(2{\bf e}_1 + 2{\bf e}_2) = 2x_1+2x_2\\
{\bf S}_b &= {\bf X}^T({\bf V}_{2b}{\bf M}_{2b}+{\bf V}_2'{\bf M}_2') = {\bf X}^T( {\bf e}_5, {\bf e}_4 + {\bf e}_6)  = (x_5, x_4+x_6) \label{eq:sb} \\
{\bf S}_c & = ({\bf X}^T{\bf V}_{1c}, {\bf X}^T{\bf V}_{2c}) = {\bf X}^T{\bf e}_7 = x_7
\end{align}
Note that ${\bf S}_b$ in (\ref{eq:sb}) is slightly different (in fact invertible) from that in (\ref{eq:sb1}) because here the invertible matrix ${\bf M}_{1b}$ is absorbed in ${\bf M}_{2b}$.
}

Let us verify that each user can recover their desired message from ${\bf S}$ and their own side-information. 
\begin{align}
{\bf S}_{a}-{\bf W}_1'{\bf Q}_1'+{\bf W}_1'{\bf P}_1'&={\bf X}^T({\bf V}_1'{\bf Q}_1'+{\bf V}_2'{\bf P}_2'-{\bf V}_1'{\bf Q}_1'+{\bf V}_1'{\bf P}_1')={\bf X}^T{\bf V}_{1a}={\bf W}_{1a}\label{eq:recw1a}\\
{\bf S}_{b}+{\bf W}_1'{\bf M}_1'&={\bf X}^T({\bf V}_{2b}{\bf M}_{2b}+{\bf V}_2'{\bf M}_2'+{\bf V}_1'{\bf M}_1')={\bf X}^T{\bf V}_{1b}={\bf W}_{1b}\label{eq:recw1b}\\
{\bf S}_{a}-{\bf W}_2'{\bf P}_2'+{\bf W}_2'{\bf Q}_2'&={\bf X}^T({\bf V}_1'{\bf Q}_1'+{\bf V}_2'{\bf P}_2'-{\bf V}_2'{\bf P}_2'+{\bf V}_2'{\bf Q}_2')={\bf X}^T[{\bf V}_{2a}, {\bf 0}]=[{\bf W}_{2a},{\bf 0}]\label{eq:recw2a}\\
({\bf S}_{b}-{\bf W}_2'{\bf M}_2'){\bf M}_{2b}^{-1}&={\bf X}^T({\bf V}_{2b}{\bf M}_{2b}+{\bf V}_2'{\bf M}_2'-{\bf V}_2'{\bf M}_2'){\bf M}_{2b}^{-1}={\bf X}^T{\bf V}_{2b}={\bf W}_{2b}\label{eq:recw2b}\\
{\bf S}_c&=({\bf W}_{1c}, {\bf W}_{2c})
\end{align}
Thus, User $1$ is able to recover ${\bf W}_{1a}$ from $({\bf S}_a,{\bf W}_1')$ according to \eqref{eq:recw1a}, ${\bf W}_{1b}$ from $({\bf S}_b, {\bf W}_1')$ according to \eqref{eq:recw1b}, and ${\bf W}_{1c}$ directly from ${\bf S}_{c}$. Similarly, User $2$ is able to recover ${\bf W}_{2a}$ from $({\bf S}_a,{\bf W}_2')$ according to \eqref{eq:recw2a}, ${\bf W}_{2b}$ from $({\bf S}_b, {\bf W}_2')$ according to \eqref{eq:recw2b}, and ${\bf W}_{2c}$ directly from ${\bf S}_{c}$. 

Finally, note that $H({\bf S})\leq n_{1a}+n_{1b}+n_{1c}+n_{2c}=H({\bf W}_1)+H({\bf W}_2\mid {\bf W}_1', {\bf W}_2', {\bf W}_1)=H({\bf W}_1\mid {\bf W}_1')+H({\bf W}_2\mid {\bf W}_1', {\bf W}_2', {\bf W}_1)$ which matches the converse. 
 \hfill\QED

\section{Proof of Lemma \ref{lemma:extra}} \label{sec:exproof}
Define the optimal normalized broadcast cost as
\begin{align}
H^* \triangleq \frac{H(w_1, w_2)}{C_\cb} = \inf \frac{H(S)}{L}.\label{eq:bcost}
\end{align}
Note that the infimum is over all feasible $S$ subject to (\ref{eq:dec1}), (\ref{eq:dec2}), (\ref{eq:structure}) and $L \in \mathbb{N}$. Now proving $C_{\cb_2} = 4/(4-\log_2(3))$ is equivalent to proving that $H^* = 4-\log_2(3)$.
To show that $H^*=4-\log_2(3)$\mbox{ bits}, we first prove a converse bound that shows $H^* \geq 4 - \log_2(3)$ in Section \ref{sec:excon}, and then an achievable scheme that shows $H^* \leq 4 - \log_2(3)$  in Section \ref{sec:exach}.

\subsection{Converse: $H^* \geq 4 - \log_2(3)$ bits}\label{sec:excon}
Let us start with a key observation, stated in the following lemma.
\begin{lemma}\label{lemma:ex}
Any achievable scheme for CB$_2$ with block length $L$, i.e., any $P_{S\mid W_1,W_2,W_1',W_2'}$ that satisfies \eqref{eq:dec1}-\eqref{eq:structure} for the $P_{w_1, w_2, w_1',w_2'}$ specified by CB$_2$, must have $H(W_1', W_2' | S) \leq L \log_2(3)$.
\end{lemma}

{\it Proof:} Denote the decoding functions of User 1 and User 2 by $\mathcal{F}_{W_1'}$ and $\mathcal{G}_{W_2'}$, respectively. The subscripts indicate that the decoding functions depend on the side-information available to each user. Because we require zero-error decoding, we must have,
\begin{align}
 &&\mathcal{F}_{W_1'}(S) = W_1,&& \mathcal{G}_{W_2'}(S) = W_2 \\
&\Leftrightarrow& \left[\mathcal{F}_{W_1'}({S})\right]_{l} = W_1(l),&& \left[\mathcal{G}_{W_2'}({S})\right]_{l} = W_2(l), &&\forall l \in [L] \label{eq:ll1}
\end{align}
where for a length $L$ sequence $A$, $\left[A\right]_l$ denotes the $l$-th symbol of $A$.

From (\ref{eq:ch1}), we note the following relationship. For any $l \in [L]$,
\begin{eqnarray} 
W_1'(l) = 0 &\Rightarrow& W_2(l) = (W_1(l) + W_2'(l)) \mod 4 \label{eq:ll2}\\
W_1'(l) = 1 &\Rightarrow& W_2(l) = (W_1(l) + 3 - W_2'(l)) \mod 4 \label{eq:ll3}
\end{eqnarray}
We now show that conditioned on any realization of $S$, and for each index $l\in[L]$, there are only three possible values for the tuple $(W_1'(l), W_2'(l))$. Here is a proof by contradiction. Suppose on the contrary that there exists some realization $S^*$ of $S$ and some index $l^* \in [L]$ such that $(W_1'(l^*), W_2'(l^*))$ can take all $4$ values in the set $\{(0,0), (0,1), (1,0), (1,1)\}$. In particular, let $A_1, A_2$ be the realizations of the length $L$ sequence $W_1'$ and $B_1, B_2$ be the realizations of the length $L$ sequence $W_2'$ such that $(A_1(l^*), A_2(l^*)) = (0,1)$ and $(B_1(l^*), B_2(l^*)) = (0,1)$. From (\ref{eq:ll1}), (\ref{eq:ll2}), (\ref{eq:ll3}), we have
\begin{eqnarray}
W_1' = A_1, W_2' = B_1 &\Rightarrow& W_1'(l^*) = 0, W_2'(l^*) = 0  \\
&\overset{(\ref{eq:ll2})}{\Rightarrow}& W_2(l^*) = (W_1(l^*) + 0) \mod 4 \\
&\overset{(\ref{eq:ll1})}{\Rightarrow}& \left[\mathcal{G}_{B_1}({S^*})\right]_{l^*} = ( \left[\mathcal{F}_{A_1}({S^*})\right]_{l^*} + 0) \mod 4 \label{eq:ll4}\\
\mbox{Similarly}, ~W_1' = A_1, W_2' = B_2 &\Rightarrow& \left[\mathcal{G}_{B_2}({S^*})\right]_{l^*} = ( \left[\mathcal{F}_{A_1}({S^*})\right]_{l^*} + 1) \mod 4 \label{eq:ll5}\\
W_1' = A_2, W_2' = B_1 &\Rightarrow& \left[\mathcal{G}_{B_1}({S^*})\right]_{l^*} = ( \left[\mathcal{F}_{A_2}({S^*})\right]_{l^*} + 3 - 0) \mod 4 \label{eq:ll6}\\ 
W_1' = A_2, W_2' = B_2 &\Rightarrow& \left[\mathcal{G}_{B_2}({S^*})\right]_{l^*} = ( \left[\mathcal{F}_{A_2}({S^*})\right]_{l^*} + 3 - 1) \mod 4 \label{eq:ll7}
\end{eqnarray}
Note that (\ref{eq:ll4}) - (\ref{eq:ll5}) - (\ref{eq:ll6}) + (\ref{eq:ll7}) gives us $0 = -2 \mod 4$, which is a contradiction. Thus, we have shown that given any realization of $S$, there are at most $3^L$ possible realizations of $(W_1', W_2')$. Using the fact that the uniform distribution maximizes entropy, $H(W_1', W_2'\mid S)\leq L\log_2(3)$ and  Lemma \ref{lemma:ex} is proved.

\hfill \QED

Equipped with Lemma \ref{lemma:ex}, the converse proof is immediate. Let us expand $H(W_1', W_2', S)$ in two ways. On the one hand, 
\begin{eqnarray}
H(W_1', W_2', S) &=& H(W_1', W_2') + H(S|W_1', W_2') \\
&=& 2L + H(S, W_1, W_2|W_1', W_2') \label{eq:ee1}\\
&\geq&  2L + H(W_1, W_2|W_1', W_2') \\
&=& 4L \label{eq:ef1}
\end{eqnarray}
where (\ref{eq:ee1}) follows from the decoding constraints, i.e., from $S, W_1', W_2'$, we can decode $W_1, W_2$ with no error.
On the other hand,
\begin{eqnarray}
H(W_1', W_2', S) &=& H(S) + H(W_1', W_2'|S) \\
&\leq& H(S) + L\log_2 (3) \label{eq:ee2}
\end{eqnarray}
as shown in Lemma \ref{lemma:ex}. Combining (\ref{eq:ef1}) and (\ref{eq:ee2}), we have
\begin{align}
\forall L\in\mathbb{N},& &H(S) + L \log_2 (3) &\geq 4L \\
\Rightarrow&& \frac{H(S)}{L}&\geq 4-\log_2(3)\\
\Rightarrow && H^*=\inf \frac{H(S)}{L}&\geq 4-\log_2(3).
\end{align}
and the proof of the converse bound $H^*\geq 4-\log_2(3)$ is complete.

\subsection{Achievability: $H^*\leq 4-\log_2(3)$ bits}\label{sec:exach}
Based on the alphabet, the set of possible values of $(w_1',w_2')$ is $\mathcal{W}_1'\times\mathcal{W}_2'=\{(0,0), (0,1), (1,0), (1,1)\}$. Note that $|\mathcal{W}_1'\times\mathcal{W}_2'|=4$. Consider an arbitrary sequence of subsets $\mathcal{W}(l)\subset \mathcal{W}_1'\times\mathcal{W}_2'$ such that $|\mathcal{W}(l)|=3$. First we show that if $(W_1'(l), W_2'(l))$ tuples are restricted to take values  in $\mathcal{W}(l)$, then sending $2L$ bits is sufficient to satisfy both users' demands. This result is stated in the following lemma.

\begin{lemma}\label{lemma:3minstr}
For any $L\in\mathbb{N}$, if for all $l \in [L]$, the tuple $(W_1'(l), W_2'(l))\in\mathcal{W}(l)\subset\mathcal{W}_1'\times\mathcal{W}_2'$, $|\mathcal{W}(l)|=3$, and the sequence $\mathcal{W}(l), l\in[L]$ is already known to the users, then broadcasting $2L$ bits is sufficient to satisfy both users' demands.
\end{lemma}

{\it Proof:} We have $4$ cases for $\mathcal{W}(l)$ as listed below.
\begin{enumerate}
\item $\mathcal{W}= \{(0,0), (0,1), (1,0)\}$.
In this case, the relationship between $W_2(l)$ and $W_1(l)$ can be described as $W_2(l) = ( W_1(l) + 3W_1'(l) + W_2'(l)) \mod 4$ such that transmitting $S(l) = (W_1(l) + 3W_1'(l)) \mod 4$ is sufficient to satisfy both users' demands. User 1 simply subtracts $3W_1'(l)$ (modulo $4$) to get $W_1(l)$, and User 2 adds $W_2'(l)$ (modulo $4$) to get $W_2(l)$. 
\item $\mathcal{W}= \{(0,0), (0,1), (1,1)\}$.
Here we have $W_2(l) = ( W_1(l) + W_1'(l) + W_2'(l)) \mod 4$ and set $S(l) = (W_1(l) + W_1'(l)) \mod 4$. User 1 simply subtracts $W_1'(l)$ to get $W_1(l)$, and User 2 adds $W_2'(l)$ to get $W_2(l)$, all modulo $4$.
\item $\mathcal{W}= \{(0,0), (1,0), (1,1)\}$.
Here we have $W_2(l) = ( W_1(l) + 3W_1'(l) - W_2'(l)) \mod 4$ and we choose to send $S(l) = (W_1(l) + 3W_1'(l)) \mod 4$. User 1  subtracts $3W_1'(l)$ from $S(l)$ to get $W_1(l)$, and User 2 subtracts $W_2'(l)$ from $S(l)$ to get $W_2(l)$, all modulo $4$. 
\item $\mathcal{W}= \{(0,1), (1,0), (1,1)\}$.
Here we have $W_2(l) = ( W_1(l) + W_1'(l) + W_2'(l) + 2) \mod 4$ and we set $S(l) = (W_1(l) + W_1'(l)) \mod 4$. User 1  subtracts $W_1'(l)$ from $S(l)$ to get $W_1(l)$, and User 2 adds $W_2'(l)+2$ from $S(l)$ to get $W_2(l)$, all modulo $4$. 
\end{enumerate}
Note that in every case, for each $l \in [L]$, $S(l)$ is a number modulo $4$ which is represented by $2$ bits, so broadcasting $2L$ bits is sufficient overall. The proof of Lemma \ref{lemma:3minstr} is thus complete.
\hfill\QED

The key to the achievable scheme is to send $\mathcal{W}(l)$ to the users, in addition to the $2L$ bits that are needed once $\mathcal{W}(l)$ is known to both users.  To describe $\mathcal{W}(l)$ it suffices to describe its complement, i.e., $(\mathcal{W}_1'\times \mathcal{W}_2')\backslash\mathcal{W}(l)$. Equivalently, we wish to describe to the users $1$ element of $\mathcal{W}_1'\times\mathcal{W}_2'$ which is not the actual realization of $(W_1'(l), W_2'(l))$ tuple  so that the users know that the actual realization is among the $3$ remaining values. Since there are $3$ values that are not the actual realization, we have $3$ choices for what to send for each $l \in [L]$. Overall, we have $3^L$ choices for $(W'_1, W'_2)$ tuples that do not match the actual realization for any $l\in[L]$. We next show that conveying one of these $3^L$ possibilities (out of the total $4^L$ possibilities) requires $(2 - \log_2 (3))L + o(L)$ bits with probability  of error $\epsilon \rightarrow 0$ as $L \rightarrow \infty$. 
This result is stated in the following lemma with general parameters, which will be used again in the proof of Theorem \ref{thm:str}.

\begin{lemma}\label{lemma:bin}
Suppose there is  a set of $n_1^L$ tuples known to a transmitter and receiver, out of which an arbitrary subset of $n_2^L$ tuples are designated acceptable, $n_1, n_2 \in \mathbb{N}, n_2 < n_1$. The acceptable tuples are known only to the transmitter, and the goal is for the transmitter to communicate any one of these acceptable tuples to the receiver. Then there exists an $\epsilon$-error scheme that allows the transmitter to accomplish this task by sending only $(\log_2 (n_1) - \log_2 (n_2))L + o(L)$ bits to the receiver.
\end{lemma}

The detailed proof of Lemma \ref{lemma:bin} is deferred to Section \ref{sec:binproof}. Let us present an outline of the proof here. The scheme is based on random binning. Throw the $n_1^L$ tuples uniformly into roughly $n_2^L$ bins. 
Pick bin 1. Find an acceptable tuple in bin 1 and send its index. Because there are $n_2^L$ bins and $n_2^L$ acceptable tuples, an $\epsilon$ change in the exponents will guarantee that each bin will typically get at least one acceptable tuple with high probability. Specifying the index of the acceptable tuple will take $\log_2(n_1^L/n_2^L) = (\log_2 (n_1) - \log_2 (n_2))L$ bits because each bin contains approximately $n_1^L/n_2^L$ tuples.

Finally, let us  summarize the overall achievable scheme which requires a minor adjustment to make it a zero-error scheme. For each realization of $(W_1, W_2, W_1', W_2')$, we use the scheme from Lemma \ref{lemma:bin} to find and specify one acceptable $(W'_1, W'_2)$ tuple, i.e., a tuple that does not match the actual realization of $(W_1'(l),W_2'(l)$ for any $l\in[L]$ to both users. With probability $1-\epsilon$, an acceptable $(W'_1, W'_2)$ tuple is found and specified, and then we use the scheme from Lemma \ref{lemma:3minstr} so that each user decodes the desired message. The total number of bits broadcast in this case is $(2 - \log_2 (3))L + o(L) + 2L$. With probability $\epsilon$, we do not find an acceptable $(W'_1, W'_2)$ tuple. In this case, we directly send $(W_1, W_2)$, and the number of bits broadcast is $8L$ bits. Therefore, the average number of bits broadcast to the users is
\begin{eqnarray}
{ (1-\epsilon) \times [(4 - \log_2(3))L + o(L)] + \epsilon \times 8L }+1
\end{eqnarray}
where $1$ extra bit is used to specify if an acceptable $(W_1',W_2')$ tuple is found. This implies that 
\begin{align}
H(S) &\leq { (1-\epsilon) \times [(4 - \log_2(3))L + o(L)] + \epsilon \times 8L }+1\\
\Rightarrow H^*&=\inf_{} \frac{H(S)}{L}\leq 4-\log_2(3).
\end{align}
 The achievability proof, i.e., the proof of the bound $H^*\leq 4-\log_2(3)$ bits, is thus complete.\hfill\QED
 
 Combining the converse and achievability proofs we have shown that $H^*=4-\log_2(3)$ bits, which implies that $C_{\cb_2}=\frac{H(w_1,w_2)}{H^*}=\frac{4}{4-\log_2(3)}$ by definition.

\subsection{Proof of Lemma \ref{lemma:bin}}\label{sec:binproof}
Fix $L \in \mathbb{N}$ and $\delta = \frac{1}{\sqrt{L}}$ such that $L(1-\delta)$ is an integer. We have $n_1^L$ tuples and $n_2^{L(1-\delta)}$ bins. For each tuple, choose a bin index independently and uniformly over $[n_2^{L(1-\delta)}]$. Denote the bin index of the $i$-th tuple by $X_i, i \in [n_1^L]$, so $X_i$ is uniformly distributed over $[n_2^{L(1-\delta)}]$.

The number of tuples with bin index 1 is $T_1 = \sum_{i\in[n_1^L]} \mathbbm{1} (X_i = 1)$. Its expected value and variance are computed as follows.
\begin{eqnarray}
\mu_1 &=& \mathbb{E}\left[\sum_{i\in[n_1^L]} \mathbbm{1} (X_i = 1) \right] = \sum_{i\in[n_1^L]} \mathbb{E}\left[ \mathbbm{1} (X_i = 1) \right] = \frac{n_1^L}{n_2^{ L(1-\delta) }} \\
\sigma_1^2 &=& \mathbb{E}\left[\left(\sum_{i\in[n_1^L]} \mathbbm{1} (X_i = 1) \right)^2\right] - \mu_1^2 
=  \mathbb{E}\left[\left(\sum_{i\in[n_1^L]} \mathbbm{1} (X_i = 1) \right) \left(\sum_{j\in[n_1^L]} \mathbbm{1} (X_j = 1) \right) \right] - \mu_1^2 \notag \\
&=& \sum_{i\in[n_1^L]} \mathbb{E}\left[\left( \mathbbm{1} (X_i = 1) \right)^2 \right] + \sum_{i\neq j, i,j\in[n_1^L]} \mathbb{E}\left[ \mathbbm{1}(X_i=1)\mathbbm(X_j=1) \right] - \mu_1^2 \\ 
&=& \frac{n_1^L}{n_2^{ L(1-\delta) }} +  \frac{n_1^{2L} - n_1^L}{n_2^{2 L(1-\delta) }} -  \frac{n_1^{2L}}{n_2^{2 L(1-\delta) }} = n_1^L \left(\frac{1}{n_2^{ L(1-\delta) }} - \frac{1}{n_2^{2 L(1-\delta) }}  \right)
\end{eqnarray}
From Chebyshev's inequality, we have
\begin{eqnarray}
\Pr(T_1 \geq (1+\delta)\mu_1) &\leq& \frac{\sigma_1^2}{\delta^2 \mu_1^2} = \frac{n_1^L \left(\frac{1}{n_2^{ L(1-\delta) }} - \frac{1}{n_2^{2 L(1-\delta) }}  \right)}{\delta^2 \frac{n_1^{2L}}{n_2^{2 L(1-\delta) }}} = \frac{ n_2^{ L(1-\delta) } - 1}{\delta^2 n_1^L}
\end{eqnarray}
Therefore, for any small constant $\epsilon$, we can find a sufficiently large $L$ such that 
\begin{eqnarray}
\Pr(T_1 \geq (1+\delta)\mu_1) \leq \epsilon/2 \label{eq:pp1}
\end{eqnarray}
Consider any $n_2^L$ acceptable tuples. Denote the bin index for the $i$-th acceptable tuple by $Y_i, i \in [n_2^L]$, and $Y_i$ is also uniform over $[n_2^{L(1-\delta)}]$. We similarly consider the number of acceptable tuples with bin index 1, denoted as $T_2 = \sum_{i\in[n_2^L]} \mathbbm{1} (Y_i = 1)$.
\begin{eqnarray}
\mu_2 &=& \mathbb{E}\left[ T_2 \right] = n_2^{L\delta}, \sigma_2^2 = \mathbb{E}\left[ T_2^2 \right] - \mu_2^2 = n_2^{L\delta}(1-n_2^{-L(1-\delta)}) \\
\Pr(T_2 = 0) &\leq& \Pr(| T_2 - \mu_2| \geq \delta \mu_2) \leq \frac{1-n_2^{-L(1-\delta)}}{\delta^2 n_2^{L\delta}}\leq \epsilon/2 \label{eq:pp2}
\end{eqnarray}

The coding scheme works as follows. When the number of tuples in bin 1, i.e.,  $T_1 \geq (1+\delta)\mu_1$, declare an error. If there is no acceptable tuple in bin 1 ($T_2 = 0$), declare an error. Otherwise, we send the index of any acceptable tuple. From (\ref{eq:pp1}), (\ref{eq:pp2}) and the union bound, the error probability is no larger than $\epsilon/2 + \epsilon/2 = \epsilon$, which can be made arbitrarily small by picking a sufficiently large $L$.

Finally, we compute the number of bits used. Note that $\delta = 1/\sqrt{L}$.
\begin{eqnarray}
\log_2 \left((1+\delta) \mu_1\right) &=&  \log_2 (1+\delta) + \log_2\left(\frac{n_1^L}{n_2^L n_2^{-L\delta}}\right) =  L \log_2 \left(\frac{n_1}{n_2}\right)  + \sqrt{L} \log_2 (n_2) + \log_2 (1+\frac{1}{\sqrt{L}}) \notag\\
&=& L (\log_2 (n_1) - \log_2 (n_2) ) + o(L)
\end{eqnarray}
Therefore, the number of bits used matches that in the lemma. The proof of Lemma \ref{lemma:bin} is complete.

\hfill\QED

\section{Proof of Theorem \ref{thm:str}} \label{sec:str}
For the proof, it will be less cumbersome to work with the optimal normalized broadcast cost as defined in \eqref{eq:bcost}. Specifically, we
first prove that $\log_2(m) \mbox{ bits} \leq H^*\leq \log_2(m) + \log _2(m_1m_2) - \log_2(m_1+m_2-1) \mbox{ bits}$ in Section \ref{sec:bound}. Then we show that the upper extreme is tight for minimally structured settings in Section \ref{sec:min}, and that the lower extreme is tight if the setting is maximally structured in Section \ref{sec:max}.

\subsection{$\log_2(m) \mbox{ bits} \leq H^*\leq \log_2(m) + \log _2(m_1m_2) - \log_2(m_1+m_2-1) \mbox{ bits}$}\label{sec:bound}
The lower bound, $H^* \geq \log_2(m)$ bits follows immediately from Theorem \ref{thm:converse}. The bound is quite obvious, as $H^* \geq H(w_1|w_1') = \log_2(m)$. The remainder of this section is aimed at proving the upper bound, $H^*\leq \log_2(m) + \log _2(m_1m_2) - \log_2(m_1+m_2-1) \mbox{ bits}$. We will construct an achievable scheme that  works for all settings of matching computation broadcast. To this end, let us introduce some definitions along with illustrative examples. Without loss of generality, we will assume $m_1 \geq m_2$.

\begin{definition}[Standard Form, $\bullet$-Set and $\circ$-Set] 
Let us attach a label to each element $(a_i,b_j), i\in[m_1], j\in[m_2]$ of the $\Pi$ matrix as follows. The $(a_i, b_j)$ element is labelled with $\bullet$ if $b_j=1$ or if $b_j=a_i+1$. Otherwise, label it with $\circ$.
We will refer to this labelling of $\Pi$ as the standard form. The set of $(a_i,b_j)$ with label $\bullet$ is called the $\bullet$-set and the set of $(a_i,b_j)$ with label $\circ$ is called the $\circ$-set. Note that the cardinality of the $\bullet$-set is $m_1+m_2-1$ and the $\circ$-set is the complement of the $\bullet$-set.
\end{definition}

For example, when $m_1=3, m_2=2$, the standard form of $\Pi$, $\bullet$-set and $\circ$-set are shown below.
\begin{eqnarray*}
\begin{array}{c|c|c|}
&w_2'=1&w_2'=2\\ \hline
w_1'=1& \bullet ~\pi_{1,1} &\bullet ~ \pi_{1,2}\\ \hline
w_1'=2& \bullet ~\pi_{2,1}  &\circ~ \pi_{2,2}  \\ \hline
w_1'=3& \bullet ~\pi_{3,1}  &\circ ~\pi_{3,2}  \\ \hline
\end{array} ~~\mbox{ (standard form)}~~
\begin{array}{l}
\bullet-\mbox{set}: \{(1,1), (2,1), (3,1), (1,2)\}\\
\circ-\mbox{set}: \{(2,2), (3,2)\}
\end{array}
\end{eqnarray*}

\begin{definition}[Translation]
Consider any cyclic shift of the rows and/or columns of $\Pi$ labelled in standard form, i.e., $\forall i\in[m_1]$, the $i$-th row is shifted to the $((i+z_1)\mod m_1)$-th row and $\forall j\in[m_2]$, the $j$-th column is shifted to the $((j+z_2) \mod m_2)$-th column, $i, z_1\in [m_1], j, z_2\in [m_2]$. The resulting $\bullet$-set and $\circ$-set are called \emph{translations}.
\end{definition}

For example, when $m_1=3, m_2=2$, all possible translations of the $\bullet$-set and the $\circ$-set are shown below.
\begin{align*}
\begin{array}{|c|c|}\hline
  \bullet & \bullet\\ \hline
 \bullet &\circ  \\ \hline
 \bullet &\circ  \\ \hline
\end{array}
&&
\begin{array}{|c|c|}\hline
  \bullet &\circ\\ \hline
 \bullet &\circ  \\ \hline
 \bullet & \bullet \\ \hline
\end{array}
&&
\begin{array}{|c|c|}\hline
  \bullet&\circ \\ \hline
 \bullet & \bullet \\ \hline
 \bullet &\circ  \\ \hline
\end{array}
&&
\begin{array}{|c|c|}\hline
  \bullet & \bullet\\ \hline
 \circ & \bullet \\ \hline
 \circ & \bullet  \\ \hline
\end{array}
&&
\begin{array}{|c|c|}\hline
 \circ & \bullet \\ \hline
 \circ& \bullet  \\ \hline
\bullet &\bullet  \\ \hline
\end{array}
&&
\begin{array}{|c|c|}\hline
 \circ  &\bullet \\ \hline
 \bullet & \bullet  \\ \hline
 \circ  &\bullet  \\ \hline
\end{array}
\end{align*} 
where the first translation is the original standard form, and the second translation is obtained by setting $z_1 = 2, z_2 = 2 = 0 \mod 2$ (rows are cyclicly shifted by 2 and columns are not shifted).

Following the notion in geometry, translation refers to a function that moves an object without rotating or flipping it. 
Intuitively, we may think of it as replicating the standard form grid pattern infinitely in space, and choosing any contiguous $m_1\times m_2$ block from that infinite grid. Such a block is a translation.

For our achievable scheme, we will only consider the $\bullet$-sets and $\circ$-sets that can be obtained by translations of the standard form. Such $\bullet$-sets and $\circ$-sets are called regular $\bullet$-sets and regular $\circ$-sets, respectively.
The importance of regular $\bullet$-sets is highlighted in the following lemma, where we show that if $(w_1', w_2')$ can only take values from a regular $\bullet$-set, then sending $\log_2(m)$ bits per symbol is sufficient to satisfy both users' demands. Essentially, the following lemma generalizes Lemma \ref{lemma:3minstr}.

\begin{lemma}\label{lemma:minstr}
For any $L\in\mathbb{N}$, if for all $l\in[L]$, the tuple $(W_1'(l), W_2'(l))\in\mathcal{W}(l)\subset[m_1]\times[m_2]$, each $\mathcal{W}(l)$ is a regular $\bullet$-set, and the sequence $\mathcal{W}(l), l\in[L]$ is already known to the users, then broadcasting $L\log_2(m)$ bits is sufficient to satisfy both users' demands.
\end{lemma}

{\it Proof:} For any $L$, consider an arbitrary regular $\bullet$-set with cyclic shifts $z_1, z_2$ so that the $\bullet$-set contains the following elements.
\begin{eqnarray}
\mbox{$\bullet$-set} &=& \{(1+z_1,1+z_2), (2+z_1,1+z_2), \cdots, (m_1+z_1,1+z_2), \notag\\
&&~(1+z_1, 2+z_2), (2+z_1,3+z_2), \cdots, (m_2-1+z_1, m_2+z_2)\}
\end{eqnarray} 
where for an element $(a_i, b_j) \in \mbox{$\bullet$-set}$, $a_i$ is interpreted modulo $m_1$ and $b_j$ is interpreted modulo $m_2$. 

We show that there exist $m_1+m_2$ permutations $\delta_1, \cdots, \delta_{m_1}, \gamma_1, \cdots, \gamma_{m_2}$ over $[m]$ such that the following equation holds.
\begin{eqnarray}
\gamma_{w_2'} \delta_{w_1'} = \pi_{w_1', w_2'}, \forall (w_1', w_2') \in \mbox{$\bullet$-set} \label{eq:per}
\end{eqnarray}
Such $\delta_{i}, \gamma_{j}, i\in[m_1], j\in[m_2]$ are chosen as follows.
\begin{eqnarray}
&& \mbox{Choose $\gamma_{1+z_2}$ to be an arbitrary permutation, say identity}.\notag\\
&& \mbox{Set}~\delta_{1+z_1} = \gamma_{1+z_2}^{-1} \pi_{1+z_1, 1+z_2}~\mbox{such that}~ \gamma_{1+z_2} \delta_{1+z_1} = \pi_{1+z_1, 1+z_2} \notag\\
&& \mbox{Set}~\delta_{2+z_1} = \gamma_{1+z_2}^{-1} \pi_{2+z_1, 1+z_2}~\mbox{such that}~ \gamma_{1+z_2} \delta_{2+z_1} = \pi_{2+z_1, 1+z_2} \notag\\
&& ~~~~~~~~~~~~~~~~~~~~~~~~~~~~~~~~~~~~\vdots \label{eq:setper} \\
&& \mbox{Set}~\delta_{m_1+z_1} = \gamma_{1+z_2}^{-1} \pi_{m_1+z_1, 1+z_2}~\mbox{such that}~ \gamma_{1+z_2} \delta_{m_1+z_1} = \pi_{m_1+z_1, 1+z_2} \notag\\
&& \mbox{Set}~\gamma_{2+z_2} = \pi_{1+z_1, 2+z_2}\delta_{1+z_1}^{-1}~\mbox{such that}~ \gamma_{2+z_2} \delta_{1+z_1} = \pi_{1+z_1, 2+z_2} \notag\\
&& \mbox{Set}~\gamma_{3+z_2} = \pi_{2+z_1, 3+z_2}\delta_{2+z_1}^{-1}~\mbox{such that}~ \gamma_{3+z_2} \delta_{2+z_1} = \pi_{2+z_1, 3+z_2} \notag\\
&& ~~~~~~~~~~~~~~~~~~~~~~~~~~~~~~~~~~~~\vdots \notag \\
&& \mbox{Set}~\gamma_{m_2+z_2} = \pi_{m_2-1+z_1, m_2+z_2}\delta_{m_2-1+z_1}^{-1}~\mbox{such that}~ \gamma_{m_2+z_2} \delta_{m_2-1+z_1} = \pi_{m_2-1+z_1, m_2+z_2} \notag
\end{eqnarray}
where we interpret the index of $\delta_i$ modulo $m_1$ and the index of $\gamma_j$ modulo $m_2$. It is easy to verify that with the choice of $\delta_i, \gamma_j$ in (\ref{eq:setper}), (\ref{eq:per}) is satisfied. The choices of $\delta_i, \gamma_j$ for any regular $\bullet$-set are fixed and known globally. The achievable scheme now works as follows.

For any realization of $(W_1(l), W_2(l), W_1'(l), W_2'(l))$, we send $S(l) = \delta_{W_1'(l)}(W_1(l))$, which contains $\log_2 (m)$ bits. Both users decode their desired messages using the following structured decoding rule. User 1 takes the received $\delta_{W_1'(l)}(W_1(l))$ and applies the permutation $\delta_{W_1'(l)}^{-1}$ to obtain $W_1(l)$. User 2 takes the received $\delta_{W_1'(l)}(W_1(l))$ and applies the permutation $\gamma_{W_2'(l)}$ to obtain
\begin{eqnarray}
\gamma_{W_2'(l)}\delta_{W_1'(l)}(W_1(l)) \overset{(\ref{eq:per})}{=} \pi_{W_1'(l), W_2'(l)}(W_1(l)) \overset{(\ref{eq:function})}{=} W_2(l)
\end{eqnarray}
Note that $(W_1'(l), W_2'(l)) \in \bullet\mbox{-set}$.
Repeating the scheme above for all $l \in [L]$ gives us the zero-error scheme that broadcasts $L\log_2 (m)$ bits. This completes the proof of Lemma \ref{lemma:minstr}. \hfill\QED

To complete the description of the general achievable scheme we must also send some information so that for each $l\in[L]$, the users know \emph{one} regular $\bullet$-set that includes the actual realization of $(W_1'(l), W_2'(l))$, so that we can apply the scheme in Lemma \ref{lemma:minstr}. Such a regular $\bullet$-set is called \emph{acceptable}.
For example, suppose $m_1=3, m_2=2$ and the actual realization of $(W_1'(1), W_2'(1))$ is $(2,1)$. Then the acceptable regular $\bullet$-set must contain $(2,1)$, which is indicated with a shaded gray region below. So the only acceptable $\bullet$-sets are the following $4 (=m_1+m_2-1)$. 
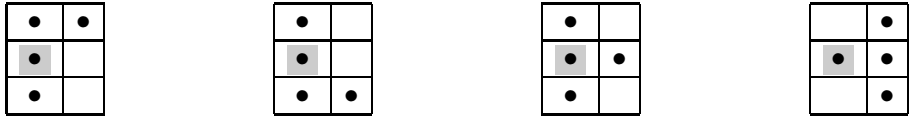
\begin{align*}
\begin{array}{|c|c|}\hline
  \bullet & \bullet\\ \hline
\colorbox{black!20!white}{$\bullet$} &  \\ \hline
 \bullet &  \\ \hline
\end{array}
&&
\begin{array}{|c|c|}\hline
  \bullet & \\ \hline
\colorbox{black!20!white}{$\bullet$} & \\ \hline
 \bullet & \bullet \\ \hline
\end{array}
&&
\begin{array}{|c|c|}\hline
  \bullet& \\ \hline
\colorbox{black!20!white}{$\bullet$}& \bullet \\ \hline
 \bullet &  \\ \hline
\end{array}
&&
\begin{array}{|c|c|}\hline
   &\bullet \\ \hline
 \colorbox{black!20!white}{$\bullet$} & \bullet  \\ \hline
   &\bullet  \\ \hline
\end{array}
\end{align*} 

In general,  for any  realization of $(W_1'(l), W_2'(l))$, let us show that there are $(m_1+m_2-1)$ acceptable regular $\bullet$-sets. This result is stated in the following lemma. 

\begin{lemma}\label{lemma:acceptable}
For any $L$ and any realization of $(W_1', W_2')$, there are $(m_1+m_2-1)^L$ acceptable regular $\bullet$-sets, out of all $(m_1m_2)^L$ regular $\bullet$-sets.
\end{lemma}

{\it Proof:} We first show that for any $l\in[L]$, there are $m_1m_2$ regular $\bullet$-sets. To this end, it suffices to show that all translations of the standard form produce distinct regular $\bullet$-sets. Consider two translated $\bullet$-sets with cyclic shifts, $(z_1, z_2), (z_1', z_2')$ such that $z_1,z_1'\in[m_1], z_2, z_2'\in[m_2],(z_1, z_2) \neq (z_1', z_2')$. Note that the $\bullet$-set in standard form contains a column where each element is labelled by $\bullet$, so if $z_2 \neq z_2'$, the two translated $\bullet$-sets are distinct (the column with all $\bullet$ is different). Now consider the case where $z_2 = z_2'$ while $z_1 \neq z_1'$. Here the two translated $\bullet$-sets are again distinct because the first row of the $\bullet$-set in standard form is distinct from all other rows and as $z_1\neq z_1'$, the first row is shifted to distinct rows. Thus in total, we have $(m_1m_2)^L$ regular sets.

Next we show that for any $l\in[L]$ and any realization $(i^*, j^*) \in [m_1]\times [m_2]$ of $(W_1'(l), W_2'(l))$, there are $m_1+m_2-1$ acceptable regular $\bullet$-sets. To see this, note that there are $m_1+m_2-1$ distinct elements labelled with a $\bullet$ in the standard form. We may shift each element to $(i^*, j^*)$, and each such shift corresponds to a distinct translation. Thus in total, we have $(m_1+m_2-1)^L$ acceptable regular $\bullet$-sets. This completes the proof of Lemma \ref{lemma:acceptable}. \hfill\QED

Combining Lemma \ref{lemma:acceptable} and Lemma \ref{lemma:bin}, we know that communicating an acceptable regular $\bullet$-set to the users requires $L(\log_2(m_1m_2) - \log_2(m_1+m_2-1))$ bits with probability $1-\epsilon$. The overall achievable scheme is described as follows. For each realization of $(W_1, W_2, W_1', W_2')$, we use the scheme from Lemma \ref{lemma:bin} to find and specify one acceptable regular $\bullet$-set. With probability $1-\epsilon$, an  acceptable regular $\bullet$-set is found, and then we use the scheme from Lemma \ref{lemma:minstr} so that each user decodes the desired message. The number of bits broadcast is $L(1-\epsilon)[\log_2(m_1m_2) - \log_2(m_1+m_2-1)+ \log_2(m)]$. For the remaining probability $\epsilon$, we directly send $(W_1, W_2)$. The number of bits broadcast is $2L\epsilon\log_2 (m)$ bits. One extra bit is used to identify the cases where $(W_1,W_2)$ are directly sent. Therefore, 
\begin{align}
H^*&=\inf_{L\in\mathbb{N}}\frac{H(S)}{L} &\leq\frac{(1-\epsilon) \times [(\log_2 (m)+\log_2 (m_1m_2)-\log_2 (m_1+m_2-1))]L+ \epsilon \times 2L\log_2 (m) +1}{L}
\end{align}
and since $\epsilon\rightarrow 0$ as $L\rightarrow\infty$, we have
\begin{align}
H^*&\leq \log_2 (m)+\log_2 (m_1m_2)-\log_2 (m_1+m_2-1).
\end{align}
The achievable scheme requires $L \rightarrow \infty$ (mainly because of the random binning operation in Lemma \ref{lemma:bin}) which is sufficient for our purpose. However, non-asymptotic schemes may also be possible. To show this, let us show an example of a finite $L$ scheme when $m_1 = 4, m_2 = 3$ ($L=1$ in fact). This example is special because $\log_2(m_1m_2) - \log_2(m_1+m_2-1)$ takes an integer value of $1$.

\subsubsection*{A non-asymptotic scheme when $m_1=4, m_2=3$}
While this scheme uses similar ideas as the asymptotic scheme, it is based on a different definition of the $\bullet$-set (not obtained by translations from standard form). Specifically for this example, the $\bullet$-set is defined as follows.
\begin{eqnarray*}
\begin{array}{c|c|c|c|}
&\gamma_1&\gamma_2&\gamma_3\\ \hline
\delta_1& \bullet ~\pi_{1,1} &\circ ~ \pi_{1,2}&\circ~\pi_{1,3}\\ \hline
\delta_2& \bullet ~\pi_{2,1}  &\bullet ~\pi_{2,2} &\circ ~\pi_{2,3} \\ \hline
\delta_3& \circ ~\pi_{3,1}  &\bullet ~\pi_{3,2} &\bullet ~\pi_{3,3} \\ \hline
\delta_4& \circ ~\pi_{4,1}  &\circ~\pi_{4,2} &\bullet ~\pi_{4,3} \\ \hline
\end{array}
\end{eqnarray*} 
Note that we label the rows and columns by the permutations $\delta_i, \gamma_j$ that we assign as follows to satisfy $\gamma_j \delta_i = \pi_{i,j}$ if $(i,j) \in \bullet$-set (following the same idea from Lemma \ref{lemma:minstr}).
\begin{eqnarray}
&& \mbox{Choose}~\gamma_1~\mbox{to be an arbitrary permutation} \notag\\
&& \mbox{Set}~\delta_1 = \gamma_1^{-1}\pi_{1,1}, \delta_2 = \gamma_1^{-1}\pi_{2,1}, \gamma_2 = \pi_{2,2} \delta_{2}^{-1} \\
&& \mbox{Set}~\delta_3 = \gamma_2^{-1}\pi_{3,2}, \gamma_3  = \pi_{3,3} \delta_3^{-1}, \delta_4 =  \gamma_3^{-1}  \pi_{4,3} \notag
\end{eqnarray}
If the users know that $(W_1'(1), W_2'(1)) \in \bullet$-set, then sending $\delta_{W_1'(1)}(W_1(1))$ ($=\log_2 (m)$ bits) is sufficient to satisfy both users's demands. After receiving $\delta_{W_1'(1)}(W_1(1))$, User 1 applies $\delta_{W_1'(1)}^{-1}$ to obtain $W_1(1)$, and User 2 applies $\gamma_{W_2'(1)}$ to obtain $\gamma_{W_2'(1)}\delta_{W_1'(1)}(W_1(1)) = \pi_{W_1'(1), W_2'(1)}(W_1(1)) = W_2(1)$. Interestingly, if $(W_1'(1), W_2'(1)) \in \circ$-set, we may assign $\delta_i, \gamma_j$ (differently) as follows such that $\gamma_j \delta_i = \pi_{i,j}$ if $(i,j) \in \circ$-set and sending $\delta_{W_1'(1)}(W_1(1))$ is sufficient to satisfy both users' demands.
\begin{eqnarray}
&& \mbox{Choose}~\gamma_1~\mbox{to be an arbitrary permutation} \notag\\
&& \mbox{Set}~\delta_3 = \gamma_1^{-1}\pi_{3,1}, \delta_4 = \gamma_1^{-1}\pi_{4,1}, \gamma_2 = \pi_{4,2} \delta_{4}^{-1} \\
&& \mbox{Set}~\delta_1 = \gamma_2^{-1}\pi_{1,2}, \gamma_3  = \pi_{1,3} \delta_1^{-1}, \delta_2 =  \gamma_3^{-1}  \pi_{2,3} \notag
\end{eqnarray}

The only remaining step is to send information so that the users know $(W_1'(1), W_2'(1))$ belong to $\bullet$-set or $\circ$-set, for which 1 bit is sufficient. The broadcast cost thus achieved is $\log_2 (m) +1 = \log_2 (m) + \log_2 (m_1m_2) - \log_2 (m_1+m_2-1)$ bits, which matches the optimal value $H^*$. 

\subsection{$H^*= \log_2(m) + \log_2(m_1m_2) - \log_2(m_1+m_2-1)$ bits if Minimally Structured}\label{sec:min}
We show that for minimally structured settings, the general achievable scheme described in Section \ref{sec:bound} is the best possible, i.e., $H^* \geq \log_2(m) + \log_2(m_1m_2) - \log_2(m_1+m_2-1)$ bits.

We start with a lemma, which is a generalization of Lemma \ref{lemma:ex}. Interpreted through the lens of induced permutations, Lemma \ref{lemma:ex} states that if the induced permutation of a length-4 cycle is a derangement, then given $S$ the set of feasible $(W_1', W_2')$ tuple values can not include all the terms of the cycle. The following lemma generalizes the same argument to cycles of any length. For simplicity, if the induced permutation of a cycle is a derangement, we say that the cycle is a derangement cycle.

\begin{lemma}\label{lemma:cycle}
For any given realization of $S$ and for any symbol index $l \in [L]$, the set of feasible values for $(W_1'(l), W_2'(l))$ contains no derangement cycle.  
\end{lemma}

{\it Proof:} The proof is by contradiction. So, let us assume that for some given realization $S^*$ of $S$, and some $l^*\in[L]$, the set of feasible values of $(W_1'(l^*), W_2'(l^*))$ contains a cycle of length $N$,
\begin{eqnarray}
(a_1,b_1)\leftrightarrow(a_2,b_2)\leftrightarrow\cdots\leftrightarrow(a_N,b_N)\leftrightarrow(a_1,b_1)
\end{eqnarray}
Thus, the feasible values for $(W_1'(l^*), W_2'(l^*))$ include all of the values in the set $\{(a_1,b_1)$, $(a_2,b_2),$ $\cdots,$ $(a_N,b_N)\}$. Let $A_1, A_2, \cdots, A_N$ denote the corresponding realizations of $W_1'$, so that we have $A_j(l^*)=a_j, j \in [N]$, and $B_1, B_2, \cdots, B_N$ denote the corresponding realizations of $W_2'$ such that $B_j(l^*)=b_j$. If $a_j=a_k$ then $A_j=A_k$, and if $b_j=b_k$ then $B_j=B_k$. Recall that $\mathcal{F}, \mathcal{G}$ denote the decoding functions of users $1$ and $2$, respectively. Based on the structure of the matching computation broadcast problem (\ref{eq:function}) and the zero-error decoding constraint (\ref{eq:dec1}), (\ref{eq:dec2}), we have
\begin{eqnarray}
 \left[\mathcal{G}_{B_1}(S^*)\right]_{l^*} &=& \pi_{a_1,b_1}\left[\mathcal{F}_{A_1}(S^*)\right]_{l^*}\nonumber\\
 \left[\mathcal{G}_{B_2}(S^*)\right]_{l^*} &=& \pi_{a_2,b_2}\left[\mathcal{F}_{A_2}(S^*)\right]_{l^*}\nonumber\\
\left[\mathcal{G}_{B_3}(S^*)\right]_{l^*} &=& \pi_{a_3,b_3}\left[\mathcal{F}_{A_3}(S^*)\right]_{l^*}\nonumber\\
\vdots\label{eq:FG}\\
 \left[\mathcal{G}_{B_N}(S^*)\right]_{l^*} &=& \pi_{a_N,b_N}\left[\mathcal{F}_{A_N}(S^*)\right]_{l^*}\nonumber
\end{eqnarray}
From the definition of a cycle, it follows that
\begin{eqnarray}
a_1=a_2 &\Rightarrow& A_1=A_2 \nonumber\\
b_2 =b_3 &\Rightarrow& B_2=B_3 \nonumber\\
a_3=a_4 &\Rightarrow& A_3 =A_4 \nonumber\\
b_4=b_5 &\Rightarrow& B_4=B_5 \nonumber\\
\vdots&\vdots&\label{eq:aabb}\\
a_{N-1}=a_N &\Rightarrow& A_{N-1}=A_N\nonumber\\
b_N=b_1 &\Rightarrow& B_N=B_1\nonumber
\end{eqnarray}
Combining (\ref{eq:FG}) and (\ref{eq:aabb}), we have
\begin{eqnarray}
 \left[\mathcal{G}_{B_N}(S^*)\right]_{l^*} &=& \pi_{a_1,b_N}\left[\mathcal{F}_{A_1}(S^*)\right]_{l^*}\nonumber\\
\left[\mathcal{G}_{B_2}(S^*)\right]_{l^*} &=& \pi_{a_1,b_2}\left[\mathcal{F}_{A_1}(S^*)\right]_{l^*}\nonumber\\
\left[\mathcal{G}_{B_2}(S^*)\right]_{l^*} &=& \pi_{a_3,b_2}\left[\mathcal{F}_{A_3}(S^*)\right]_{l^*}\nonumber\\
\vdots\label{eq:FGnew}\\
 \left[\mathcal{G}_{B_N}(\vec{s})\right]_{i^*} &=& \pi_{a_{N-1},b_N}\left[\mathcal{F}_{A_{N-1}}(S^*)\right]_{l^*}\nonumber
\end{eqnarray}
which implies that
\begin{eqnarray}
\left[\mathcal{G}_{B_N}(S^*)\right]_{l^*}&=&\pi_{a_1,b_N}\pi_{a_1,b_2}^{-1}\pi_{a_3,b_2}\cdots \pi_{a_{N-1}b_N}^{-1}\left[\mathcal{G}_{B_N}(S^*)\right]_{l^*}
\end{eqnarray}
Note that the cycle is a derangement cycle, so the induced permutation $\pi_{a_1,b_N}\pi_{a_1,b_2}^{-1}\pi_{a_3,b_2}\cdots \pi_{a_{N-1}b_N}^{-1}$ is a derangement, i.e., there is no fixed point. 

However, note that $$\left[\mathcal{G}_{B_N}(S^*)\right]_{l^*} = W_2(l^*) = \pi_{a_1,b_N}\pi_{a_1,b_2}^{-1}\pi_{a_3,b_2}\cdots \pi_{a_{N-1}b_N}^{-1}(W_2(l^*)),$$ so the decoding is incorrect.
Thus, we arrive at the contradiction which completes the proof of Lemma \ref{lemma:cycle}.\hfill \QED

Note that Lemma \ref{lemma:cycle} holds in general, e.g., it is not limited to minimally structured settings. Next, for minimally structured settings we show that if a set of values for $(W_1'(l), W_2'(l))$ contains no derangement cycle, then the cardinality of the set is no more than $m_1+m_2-1$. The intuitive reason is that a set of values for $(W_1'(l), W_2'(l))$ with more than $m_1+m_2-1$ elements over $[m_1] \times [m_2]$ must contain a cycle and every cycle is a derangement cycle for minimally structured settings. This result is stated in the following lemma.

\begin{lemma}\label{lemma:cyclemin}
For minimally structured settings, if the set $\mathcal{M} \subset [m_1] \times [m_2]$ contains no derangement cycle, then 
\begin{eqnarray}
|\mathcal{M}| \leq m_1 + m_2 - 1
\end{eqnarray}
\end{lemma}

{\it Proof:} Since every cycle for a minimally structured setting is a derangement cycle, we only need to show that $|\mathcal{M}| \leq m_1 + m_2 - 1
$ for cycle-free $\mathcal{M} \subset [m_1] \times [m_2]$.
Let the elements of $[m_1]\times[m_2]$ be mapped to the $m_1\times m_2$ table under the natural ordering. Remove any rows or columns of the table that have no elements of $\mathcal{M}$, leaving us with $m_1'\leq m_1$ rows and $m_2'\leq m_2$ columns. This cannot introduce cycles, so it suffices to show that $|\mathcal{M}|\leq m_1'+m_2'-1$, for cycle-free $\mathcal{M}\subset[m_1']\times[m_2']$. This is equivalent to the original statement of the lemma, so without loss of generality we can assume that $(m_1',m_2')=(m_1,m_2)$. Now, find a row or a column of the table that has exactly $1$ element of $\mathcal{M}$. There must exist such a row or column, because otherwise $\mathcal{M}$ contains a cycle. Eliminate this row or column, and remove the corresponding element from $\mathcal{M}$. So it now remains to show that $|\mathcal{M}|-1\leq m_1+m_2-2$, which is also equivalent to the original statement, i.e., the proof for the reduced setting implies the proof for the original setting. Continue this step, until  there remains only one row or only one column. Without loss of generality, suppose in the end we have $m_1$ rows and one column. Then we only have to show that any subset of this table cannot have more than $m_1$ elements, which is trivially true. Hence, Lemma \ref{lemma:cyclemin} is proved.\hfill \QED

The converse proof is a simple consequence of  the above two lemmas. From Lemma \ref{lemma:cycle} and Lemma \ref{lemma:cyclemin}, we know that given any realization of $S$, the number of feasible values for $(W_1', W_2')$  is no more than $(m_1+m_2-1)^L$, i.e., $H(W_1', W_2'|S) \leq L\log_2 (m_1+m_2-1)$. Then we expand $H(S, W_1', W_2)$ in two ways, similar to the proof of Lemma \ref{lemma:extra}.
\begin{eqnarray}
H(S, W_1', W_2) &=& H(W_1', W_2') + H(S|W_1', W_2') = L\log_2 (m_1m_2) + L\log_2 (m) \label{eq:summ}\\
&=& H(S) + H(W_1', W_2'|S) \leq H(S) + L\log_2 (m_1+m_2-1) \\
\Rightarrow  H(S)/L &\geq& \log_2(m) + \log_2(m_1m_2) - \log_2(m_1+m_2-1)
\end{eqnarray}
As $H^* = \inf_{} H(S)/L$, the desired bound follows and the proof of the converse bound, $H^* \geq \log_2(m) + \log_2(m_1m_2) - \log_2(m_1+m_2-1)$ bits for minimally structured settings is thus complete.

\subsection{$H^* = \log_2(m)$ bits if Maximally Structured}\label{sec:max}
We show that for maximally structured settings, the broadcast cost $\log_2(m)$ is achievable, which is a simple consequence of Lemma \ref{lemma:minstr}. Specifically, we show that although the choice of $\delta_i, \sigma_j$ in Lemma \ref{lemma:minstr} (refer to (\ref{eq:setper})) is designed to satisfy
\begin{eqnarray}
\gamma_{w_2'} \delta_{w_1'} = \pi_{w_1', w_2'} \label{eq:per2}
\end{eqnarray}
for all $(w_1', w_2')$ from only a $\bullet$-set (refer to (\ref{eq:per})), in fact it automatically satisfies \eqref{eq:per2} for all $(w_1', w_2') \in [m_1] \times [m_2]$ if the setting is maximally structured. Specifically, following (\ref{eq:setper}), we proceed as follows.
\begin{eqnarray}
&& \mbox{Choose $\gamma_1$ to be an arbitrary permutation}\notag\\ 
&& \mbox{Set}~\delta_1 = \gamma_1^{-1} \pi_{1,1},  \delta_2 = \gamma_1^{-1} \pi_{2,1}, \cdots,  \delta_{m_1} = \gamma_1^{-1} \pi_{m_1,1} \label{eq:setper2}\\
&& \mbox{Set}~\gamma_2 = \pi_{1,2} \delta_1^{-1}, \gamma_3 = \pi_{2,3} \delta_2^{-1}, \cdots, \gamma_{m_2} = \pi_{m_2-1,m_2} \delta_{m_2-1}^{-1} \notag
\end{eqnarray}
and show that (\ref{eq:per2}) is satisfied for all $(w_1', w_2') \in [m_1] \times [m_2]$ for maximally structured settings. For any $(w_1', w_2') \in [m_1] \times [m_2]$, we have a length-4 cycle $(w_1'-1,1) \leftrightarrow (w_1'-1,w_2') \leftrightarrow (w_1',w_2') \leftrightarrow (w_1', 1) \leftrightarrow (w_1'-1,1)$. As the setting is maximally structured, the induced permutation $\pi_{w_1'-1, 1} \pi_{w_1'-1, w_2'}^{-1} \pi_{w_1', w_2'} \pi_{w_1', 1}^{-1}$ is an identity. We have
\begin{eqnarray}
\mbox{Identity} &=& \pi_{w_1'-1, 1} \pi_{w_1'-1, w_2'}^{-1} \pi_{w_1', w_2'} \pi_{w_1', 1}^{-1} \\
&\overset{(\ref{eq:setper2})}{=}&\gamma_1\delta_{w_1'-1} (\gamma_{w_2'}\delta_{w_1'-1})^{-1} \pi_{w_1', w_2'} (\gamma_1\delta_{w_1'})^{-1} \\
&=& \gamma_1 \delta_{w_1'-1} \delta_{w_1'-1}^{-1} \gamma_{w_2'}^{-1} \pi_{w_1', w_2'} \delta_{w_1'}^{-1} \gamma_1^{-1} \\
\Rightarrow \gamma_{w_2'}\delta_{w_1'} &=& \pi_{w_1', w_2'}
\end{eqnarray}
so that (\ref{eq:per2}) is satisfied for all $(w_1', w_2') \in [m_1] \times [m_2]$.

The remaining description of the achievable scheme is the same as that in Lemma \ref{lemma:minstr}. For any $l\in[L]$, we send $S(l) = \delta_{W_1'(l)}(W_1(l))$, which requires $\log_2(m)$ bits. User 1 takes the received $\delta_{W_1'(l)}(W_1(l))$ and applies the permutation $\delta_{W_1'(l)}^{-1}$ to obtain $W_1(l)$. User 2 takes the received $\delta_{W_1'(l)}(W_1(l))$ and applies the permutation $\gamma_{W_2'(l)}$ to obtain
\begin{eqnarray}
\gamma_{W_2'(l)}\delta_{W_1'(l)}(W_1(l)) \overset{(\ref{eq:per2})}{=} \pi_{W_1'(l), W_2'(l)}(W_1(l)) \overset{(\ref{eq:function})}{=} W_2(l).
\end{eqnarray}
The broadcast cost thus achieved is $\log_2(m)$ bits. For maximally structured settings, we note that it suffices to set $L=1$ because there is no need to send additional information in the manner of Lemma \ref{lemma:acceptable}. The proof that $H^* = \log_2(m)$ bits for maximally structured settings, is thus complete.

\section{Conclusion}
The computation broadcast problem represents a small step towards  an understanding of the dependencies that exist across message flows and side-informations when communication networks are used for distributed computing applications. Since  linear computations are quite common, the capacity characterization for the linear computation broadcast problem is significant. The immediate question for future work is to find the capacity of linear computation broadcast for more than $2$ users. The question is particularly interesting because even the $3$ user setting appears to be non-trivial, i.e., it does not follow as a direct extension from the $2$ user case studied here. Beyond linear settings, a number of questions remain open even for $2$ users. While the general converse bound of Theorem \ref{thm:converse} uses only entropic structure, it is not known if it captures \emph{all} of the entropic structure, i.e., whether the bound is always tight for the entropic formulation of the computation broadcast problem. Another interesting problem is to use the insights from the linear and matching computation broadcast problems to construct powerful achievable schemes for general computation broadcast, even for two users. For example, is it possible to create an efficient $a,b,c$ partition of a general computation broadcast problem? If so, then the optimal solutions for $a$ and $c$ partitions are already known in the general case, which leaves us with only the $b$ partition, i.e., the minimally dependent part of the problem. The matching problems appear to be the key to the general solution of such settings. The exact capacity for matching computation broadcast problems also remains open for settings that are neither maximally structured nor minimally structured. A remarkable insight from the capacity characterization for minimally structured settings is that it is better to exploit local structure even with the additional overhead cost of identifying this local structure to both receivers (this overhead is the greatest in minimally structured settings), rather than the obvious alternative, which is to ignore the minimal structure and simply use random coding. The possibility of generalizing this intuition to broader classes of computation broadcast is worth exploring as well. Evidently, the computation broadcast problem presents a fresh opportunity to explore some of the deeper questions in information theory regarding the structure of information, in a setting that is most appealing for its simplicity --  involving only $5$ random variables: $W_1, W_1', W_2, W_2', S$.

\section*{Appendix: Proofs of Lemma \ref{lemma:obvious} and Lemma \ref{lemma:ind}}
{\it Proof of Lemma \ref{lemma:obvious}:} 
From the definition of the rank function, there exist $\mu=\mbox{rank}({\bf A})$ column vectors of the matrix ${\bf A}$ that are linearly independent. Denote the matrix formed by these vectors ${\bf A}_{sub}$. The column vectors of ${\bf A}$ are linear combinations of those of ${\bf A}_{sub}$, i.e., ${\bf X}^T {\bf A}$ are deterministic functions of ${\bf X}^T {\bf A}_{sub}$. Therefore we have
\begin{align}
H({\bf X}^T {\bf A}) = H({\bf X}^T {\bf A}_{sub})
\end{align}
It suffices now to prove that $H({\bf X}^T {\bf A}_{sub}) \leq \mu$ and $H({\bf X}^T {\bf A}_{sub}) \geq \mu$. It is trivial to see that $H({\bf X}^T {\bf A}_{sub}) \leq \mu$ because ${\bf X}^T {\bf A}_{sub}$ contains only $\mu$ elements in $\mathbb{F}_q$ so its entropy cannot be more than $\mu$ in $q$-ary units (uniform distribution maximizes entropy).
Next, we show that $H({\bf X}^T {\bf A}_{sub}) \geq \mu$. From the definition of the rank function, ${\bf A}_{sub}$ contains a square $\mu\times\mu$ invertible sub-matrix. Denote this sub-matrix as ${\bf A}_{squ}$. Without loss of generality, assume ${\bf A}_{squ}$ is formed by the first $\mu$ rows of ${\bf A}_{sub}$.
\begin{align}
H({\bf X}^T {\bf A}_{sub}) &\geq H({\bf X}^T {\bf A}_{sub} \mid x_{\mu+ 1}, \cdots, x_{m-1}, x_{m})\\
&= H([x_1, x_2, \cdots, x_\mu] {\bf A}_{squ} \mid x_{\mu + 1}, \cdots, x_{m-1}, x_{m})\\
&= H(x_1, x_2, \cdots, x_\mu \mid x_{\mu + 1}, \cdots, x_{m-1}, x_{m}) \label{eq:inv1} \\
&= \mu
\end{align}
where (\ref{eq:inv1}) follows from the fact that ${\bf A}_{squ}$ is invertible and applying invertible transformations does not change the entropy, and the last step is due to the condition that $x_1, \cdots, x_m$ are i.i.d. uniform over $\mathbb{F}_q$. This completes the proof of Lemma \ref{lemma:obvious}.

\noindent {\it Proof of Lemma \ref{lemma:ind}:} 
Lemma \ref{lemma:ind} follows immediately from Lemma \ref{lemma:obvious}. Note that 
\begin{align}
I({\bf X}^T {\bf A}; {\bf X}^T {\bf B}) &= H({\bf X}^T {\bf A}) + H({\bf X}^T {\bf B}) - H({\bf X}^T[{\bf A}, {\bf B}]) \\
&= \mbox{rank}({\bf A}) + \mbox{rank}({\bf B}) - \mbox{rank}([{\bf A}, {\bf B}]) \label{eq:rankto}
\end{align}
where we have used Lemma \ref{lemma:obvious} in the last step. Therefore $I({\bf X}^T {\bf A}; {\bf X}^T {\bf B}) = 0$ if and only if $\mbox{rank}({\bf A}) + \mbox{rank}({\bf B}) = \mbox{rank}([{\bf A}, {\bf B}])$, which is in turn equivalent to that $\mbox{span}(A)$ and $\mbox{span}(B)$ are independent subspaces. This completes the proof of Lemma \ref{lemma:ind}.

\bibliographystyle{IEEEtran}
\bibliography{Thesis}
\end{document}